\documentclass[letterpaper,twocolumn,superscriptaddress,showpacs,nofootinbib,notitlepage,accepted=2021-04-13]{quantumarticle}
\pdfoutput=1
\usepackage[utf8]{inputenc}
\usepackage[numbers,sort&compress]{natbib}

\usepackage[english]{babel}
\usepackage[T1]{fontenc}

\usepackage{graphicx,color,dsfont,xcolor,amsmath, amssymb}
\usepackage{gensymb}
\usepackage{braket}
\usepackage{hyperref}
\usepackage{bbold}

\def\onlinecite{\cite}

\definecolor{darkblue}{RGB}{0,0,127}

\def\vr{\mathbf{r}}
\def\vs{\mathbf{s}}
\def\vy{\mathrm{y}}
\def\vx{\mathrm{x}}
\def\vf{\mathrm{f}}
\def\rmfrac{\mathrm{frac}}
\def\matr{\mathbf}

\usepackage[all]{xy}

\begin{document}

\title{
Fractalizing quantum codes
}

\date{April 15, 2021}

\author{Trithep Devakul}
\affiliation{Department of Physics, Princeton University, Princeton, NJ 08540, USA}
\affiliation{Department of Physics, Massachusetts Institute of Technology, Cambridge, MA 02139, USA}
\author{Dominic J. Williamson}
\affiliation{Stanford Institute for Theoretical Physics, Stanford University, Stanford, CA 94305, USA}

\begin{abstract} 
We introduce ``fractalization'', a procedure by which spin models are  extended to higher-dimensional ``fractal'' spin models.  
This allows us to interpret type-II fracton phases, fractal symmetry-protected topological phases, and more, in terms of well understood lower-dimensional spin models. 
Fractalization is also useful for deriving new spin models and quantum codes from known ones.
We construct higher dimensional generalizations of fracton models that host extended fractal excitations.
Finally, by applying fractalization to a 2D subsystem code, we produce a family of locally generated 3D subsystem codes that are conjectured to saturate a quantum information storage tradeoff bound.
\end{abstract}

\maketitle

\section{Introduction}
Recent years have seen increasing interest in various spin models that are defined on regular lattices and nevertheless exhibit ``fractal'' properties. 
These include gapped spin liquid models in which immobile topological excitations are created at the corners of operators with fractal support, or
 spin models with symmetries that act on a fractal subsystem.

One example of the former is Haah's code~\cite{Haah}, a canonical model of type-II~\cite{vijay2016fracton} fracton topological order~\cite{nandkishorereview,pretkoreview,chamon,bravyi2011topological,haah2011local,kim20123d,yoshida2013exotic,vijayhaahfu2015,vijay2016fracton,williamsongauge,Dua2019a}.
Such 3D phases are characterized by topological quasiparticle excitations that are strictly immobile.
As quantum codes, they lack string-like logical operators, and instead have logical operators supported on a fractal subset of sites.
The fractal nature of these codes leads to promise as quantum memories~\cite{PhysRevLett.107.150504,bravyi2013quantum,brown2014quantum}.
More generally, fracton phases have received tremendous attention in a wide variety of contexts~\cite{PhysRevB.95.155133,shriyaprx,gromovprr, PhysRevLett.123.136401,PhysRevB.101.165145,PhysRevB.99.155126, PhysRevLett.120.195301,gromov2017fractional,PhysRevB.100.134113,Gromov2019,PhysRevLett.119.257202,prem2018pinch,Doshi2020,PhysRevD.96.024051,Brown2019,vijay2017generalization,Williamson2020,Bulmash2019,Prem2019,Stephen2020,Pai2019,haah2014bifurcation,shirley2017fracton,Dua2019b,Slagle2018,Slagle2018b,Wen2020,Aasen2020,wang2020nonliquid,bulmash2018generalized,Gromov2018,PretkoGauge2018,Williamson2018,Seiberg_2020,Tantivasadakarn_2020,Shirley2020,Tantivasadakarn_JW,Devakul2020,PhysRevB.97.041110}.

An example of the latter type of fractal models are the fractal Ising models~\cite{fractalSSPT}.  
These are classical spin models on the square lattice with symmetries that flip the Ising spins on a fractal subset of sites.  
These have been studied as classical codes~\cite{CAcode,yoshidaclassical,Nixon2020} for their information storage capacity and also as translation invariant models of glassiness~\cite{newmanmoore}.
In these codes, classical information is stored in the spontaneous symmetry-broken ground states of the model.
With the same set of fractal symmetries, there may also be non-trivial phases without spontaneous symmetry breaking, known as fractal symmetry-protected topological (SPT) phases~\cite{fractalSSPT,kubica2018ungauging,Devakul2018}.
An example is the cluster model~\cite{cluster} on the honeycomb lattice, which realizes a non-trivial SPT phase protected by fractal subsystem symmetries~\cite{fractalSSPT}.
In 2D, such phases have received attention as they have been proven to be useful resources for universal measurement-based quantum computation~\cite{Raussendorf_2019,devakul2018universal,stephen2019subsystem,Daniel2019}.

In this paper, we unify all these fractal spin models via a process called ``fractalization''.
Fractalization maps certain operators defined on a $D$ dimensional lattice to one on a $D+m$ dimensional lattice, taking as input a set of $m$-dimensional linear cellular automaton (LCA) rules.
In particular, the various fractal models above may all be understood as fractalized versions of simple lower-dimensional models.
A class of type-II fracton topological phases~\cite{yoshida2013exotic} in 3D, including Haah's cubic code, may be understood as fractalized 2D toric codes~\cite{kitaev}; the fractal Ising models in 2D are simply fractalized 1D Ising models; and the 2D cluster models realizing fractal SPT phases are fractalized 1D cluster models. 
We also present an interpretation of the fractalization map as a three-step process which directly relates many properties of the fractalized model to that of the original.  
For instance, we show that the ground state manifold of a fractalized commuting Hamiltonian is (non-local) unitarily related to that of stacks of the original model.

We discuss the nature of the excitations in fractalized models in relation to those of the original model, taking as examples the toric code in various dimensions.
Point-like excitations of the original model lead to immobile point-like fracton excitations in the fractalized model, whose immobility is guaranteed by the lack of string-like logical operators.
When the original model has loop-like excitations, on the other hand, 
we determine a criterion under which the fractalized model lacks any loop or string-like excitations whatsoever.  
The excitations of such models are instead ``fractalized loop'' excitations: extended deformable fractal excitations with energy cost scaling faster than linearly with the linear size.  

When applied to quantum codes, fractalization results in a higher dimensional code with information storage capabilities similar to that of decoupled stacks of the original, but with improved code distance.  
In particular, for a locally generated $[n,k,d]$ code in $D$ dimensions, where $n$ is the number of physical qubits, $k$ is the number of logical qubits, and $d$ is the code distance, the $D\rightarrow D+m$ fractalized code is a locally generated $[L^m n, L^m k, d^\prime]$ code in $D+m$ dimensions, where $d^\prime\geq d$ depending on the LCA rules and $L$ is the linear size of the system.
Fractalization therefore has a useful application in deriving new fractal codes from known lower-dimensional codes.  
By fractalizing the 4D toric code, for instance, we construct a 5D model which lacks any membrane logical operators, a direct generalization of Type-II fracton phases in 3D whose hallmark is the absence of string logical operators~\cite{vijay2016fracton}.
We also discuss in detail fractalization applied to the 2D Bacon-Shor subsystem code~\cite{bacon,poulin}, which results in the 3D fractal Bacon-Shor code.
By incorporating ideas from Bravyi~\cite{bravyi} and Yoshida~\cite{yoshidaclassical}, we are able to improve the information storage capabilities of the 3D fractal Bacon-Shor code to $[n,k,d]\sim [L^3,L^2,L^{\eta}]$, where we conjecture $\eta\rightarrow 2$ in a limit of large physical qudit dimension.
To the best of our knowledge, if our conjecture proves true this would be the first code to saturate the information storage tradeoff bound for locally generated subsystem codes in 3D~\cite{bravyi,flammia}, $k\sqrt{d}\leq \mathcal{O}(n)$.

In Sec~\ref{sec:prelim}, we review how LCA rules generically lead to fractal structures along with other necessary background.
In Sec~\ref{sec:fractalization}, we define and discuss the fractalization procedure and its interpretation in terms of a three-step process.
In Sec~\ref{sec:examples}, we go through each of the mentioned examples in detail, and discuss connections to other recent works.
In Sec~\ref{sec:fractalizingloops}, we discuss fractalized loop excitations in detail through examples, including higher dimensional toric codes. 
In Sec~\ref{sec:fsc}, we present the fractalized Bacon-Shor code and discuss its memory storage capabilities. 
Finally, concluding remarks and future directions are laid out in Sec~\ref{sec:conclusion}.

\section{Preliminaries}\label{sec:prelim}
\subsection{Fractals and linear cellular automata}\label{sec:review}
We begin by reviewing linear cellular automata (LCA) and their connection to fractals.  
Let $c_{\vr,t}\in\{0,1\}$ represent the state of cell $\vr=(r_1,\dots, r_D)\in \mathbb{Z}^D$ of a cellular automaton at time $t$.  
The LCA update rule determines how the state is determined as a linear function (modulo 2) of the state at previous time.
Throughout this paper, we are mostly concerned with first-order, local, and translation invariant LCA update rules.
That is,
\begin{equation}
c_{\vr,t} = \sum_{\vr^\prime} F_{\vr,\vr^\prime} c_{\vr^\prime,t-1} \equiv \mathbf{F}\mathbf{c}_{t-1}
\label{eq:updatevec}
\end{equation}
where the binary matrix $\mathbf{F}$ specifies the LCA rules, and all arithmetic is implied modulo 2.
First-order means that $\mathbf{c}_{t}$ only depends on the states $\mathbf{c}_{t-1}$ at time $t-1$, locality implies that $\mathbf{F}$ is only non-zero for small $|\vr-\vr^\prime|$, and translation invariance means that $F_{\vr,\vr^\prime}=f_{\vr-\vr^\prime}$ only depends on the difference $\vr-\vr^\prime$.

Starting at time $t=0$ with the state $\mathbf{c}_{0}$, the space-time trajectory at all future times is completely determined by
\begin{equation}
\mathbf{c}_{t} = \mathbf{F}^t \mathbf{c}_{0} 
\label{eq:updatevect0}
\end{equation}
The set of cells with $c_{\vr,t}=1$ will generically form a fractal subset of the $(D+1)$ dimensional space-time lattice.
This is most naturally seen by adopting a polynomial representation for LCAs.  

In the polynomial representation, the state $\mathbf{c}_t$ is represented by a $D$-variate polynomial with $\mathbb{F}_2$ coefficients,  
\begin{equation}
c_t(\vx) = \sum_\vr c_{\vr,t} \vx^{\vr}\in\mathbb{F}_2[\vx],
\end{equation}
where $\vx=\{x_i\}_{i\in[1\dots D]}$, and $\vx^{\vr} \equiv \prod_i x_i^{r_i}$ is a monomial.
Similarly, the update rule is represented by
\begin{equation}
f(\vx) = \sum_{\vr} f_\vr \vx^\vr
\end{equation}
in terms of which Eqs.~\ref{eq:updatevec} and \ref{eq:updatevect0} take the particularly simple form 
\begin{equation}
c_t(\vx) = f(\vx) c_{t-1}(\vx) = f(\vx)^t c_0(\vx).
\end{equation}
We will utilize both the vectoral representation and the polynomial representation, depending on whichever one is most convenient.

We always assume $f(\vx)$ is chosen to have a non-zero constant term and only positive powers of $x_i$, and that $f(\vx)\neq 1$ is non-trivial.
To see that this generates fractal structures, notice that for any $t=2^n$, we have, due to the properties of $\mathbb{F}_2$ polynomials, $f(\vx)^{2^n} = f(\vx^{2^n})=f(\{x_i^{2^n}\})$.
Assuming the initial state $c_0(\vx)$ only contained finitely high powers of any $x_i$, then at exponentially long times $c_t(\vx)=f(\vx^{2^n})c_0(\vx)$ describes copies of the initial state each shifted by $2^n$ for each term in $f(\vx)$.

We most often consider LCA defined on finite $(L_1,\dots,L_d)$ systems with periodic boundary conditions, which is enforced by the identification $x_i^{L_i}=1$.
An important property is the reversibility of an LCA rule for a given system size, which corresponds to the existence of $f(\vx)^{-1}$ satisfying $f(\vx)^{-1} f(\vx) = 1$, after taking into account periodic boundary conditions.
A simple family of reversible LCA is given by any polynomial $f(\vx)$ satisfying $f(1)=1$ on a system of size $L_i=2^n$.
This follows from the fact that $f(\vx)^{2^n} = f(\vx^{2^n}) = f(1)=1$, and so the inverse is given by $f(\vx)^{-1}\equiv f(\vx)^{2^n-1}$.

As an example, consider $d=1$ spatial dimensions, $f(x)=1+x$ (corresponding to $F_{\vr,\vr^\prime}=\delta_{\vr,\vr^\prime} + \delta_{\vr,\vr^\prime+1}$) and the initial state $c_0(x)=1$ (corresponding to $c_{\vr,0} = \delta_{\vr, 0}$).
Then $c_t(x)=f(x)^t$, and listing out terms for the powers of $f(x)$ we find
\begin{equation}
\begin{matrix}
f(x)^0 & = & 1 & & & & &\\
f(x)^1 & = & 1 &x& & & &\\
f(x)^2 & = & 1 & &x^2& & &\\
f(x)^3 & = & 1 &x&x^2&x^3 &&\\
f(x)^4 & = & 1 & & & &x^4&\\
f(x)^5 & = & 1 &x & & &x^4&x^5\\
f(x)^6 & = & 1 &  &x^2 & &x^4&&x^6\\
f(x)^7 & = & 1 &x &x^2&x^3&x^4&x^5&x^6&x^7\\
\end{matrix}
\label{eq:sierpinski}
\end{equation}
and so on.
Zooming out, the fractal generated is the Sierpinski triangle (Pascal's triangle mod 2).
We remark that $f(x)=1+x$ is irreversible on any system size.  
A simple example of a reversible LCA is ${f(x)=1+x+x^2}$ on sizes $L=2^n$.

\subsection{Pauli operators}

In this paper, we present a procedure by which a set of LCA rules may be used to extend a quantum CSS code into a higher dimensional fractal code.
To this end, it is useful to use a vectoral (and polynomial) representation of Pauli operators~\cite{nla.cat-vn2487773,haah2013commuting,yoshida2013exotic}. 
Let us consider qubits on the sites $\vr$ of a hypercubic lattice.
Acting on these qubits, we have Pauli operators $X_{\vr}$ and $Z_{\vr}$.
Define a map $\sigma_X$ which maps a binary vector $a_{\vr}$ to a tensor product of $X$ Pauli operator as
\begin{equation}
\sigma_X[\mathbf{a}] = \prod_\vr X_\vr^{a_\vr},
\end{equation}
and similarly $\sigma_Z$,
\begin{equation}
\sigma_Z[\mathbf{b}] = \prod_\vr Z_\vr^{b_\vr}.
\end{equation}
The commutation relation between two Pauli operators is straightforward to compute,
\begin{equation}
[[\sigma_X[\mathbf{a}]  ,\sigma_Z[\mathbf{b}] ]]
=
(-1)^{\mathbf{b}^T \mathbf{a}} 
\end{equation}
where $[[A,B]] = A^{-1} B^{-1} A B$ is the group commutator.

Using the correspondence between vectors and polynomials, $\mathbf{a} \leftrightarrow \sum_\vr a_\vr \vx^\vr$, $\sigma_{X/Z}$ may also be interpreted as a map from $\mathbb{F}_2$ polynomials to Pauli operators.
Translations are nicely expressed in this language as multiplication by a monomial:
$\sigma_X[a(\vx)]$ translated by $\vr$ is $\sigma_X[\vx^\vr a(\vx)]$.
Note that we use the same symbol $(\sigma_{X/Z})$ in both the vectoral and polynomial representation.

The commutation relation for translations of two operators can be neatly summed up by a single commutation polynomial $c(\vx)$,
\begin{equation}
[[\sigma_X[\vx^\vr a(\vx)],  \sigma_Z[\vx^{\vr^\prime}b(\vx)] ]]
=
(-1)^{c_{\vr^\prime-\vr}} 
\end{equation}
where
\begin{equation}
c(\vx) = \sum_{\vr} \vx^\vr c_\vr = a(\vx) b(\bar{\vx})
\end{equation}
and $\bar{\vx}\equiv \vx^{-1}$.
In particular, if $c(\vx)=0$, then all translations commute.

Throughout this paper, we typically consider more than one physical qubit per lattice site.  
The generalization to $N$ qubits per site, with operators $X_\vr^{(n)}$ and $Z_{\vr}^{(n)}$ for $n=1\dots N$, is straightforward:  
in the vectoral representation, $\sigma_{X/Z}$ now takes as input $N$ vectors (or a tensor), $a_{\vr,n}$, one for each of the qubits.
Similarly, in the polynomial representation, $\sigma_{X/Z}$ takes as input a vector of $N$ polynomials, $\mathbf{a}(\vx)=a_n(\vx)$.
The commutation polynomial is then given by $c(\vx) = \mathbf{a}^T(\vx) \mathbf{b}(\bar{\vx})$.

\section{Fractalization}
\label{sec:fractalization}

\subsection{Definition}\label{sec:fracdef}

We are now ready to discuss fractalization. 
In the most general case, fractalization maps a $D$-dimensional $X$ or $Z$ Pauli operator to one in $D+m$ dimensions.
We consider a hypercubic lattice with $N$ qubits per site.
As input, we take a set of $D$, $m$-dimensional LCA rules, ${\mathrm{f}(\vy)\equiv\{f_i(\vy)\}_{i\in[1\dots d]}}$, where $\vy=\{y_j\}_{j\in[1\dots m]}$.

Let $A$ be a local $X$ Pauli operator in $D$ dimensions, in the polynomial representation, 
\begin{equation}
A = \sigma_X\left[\vx^{\vr_0}\mathbf{a}(\vx)\right],
\end{equation}
where $\vx=\{x_1,\dots,x_d\}$ and we have chosen an ``anchor point'' $\vr_0$.
We uniquely specify the anchor point for this operator by maximizing each $(\vr_0)_i$ subject to the requirement that $\mathbf{a}(\vx)$ has only positive powers of $x_i$.
Locality means that $\mathbf{a}(\vx)$ contains only finitely high powers of $x_i$.

Going to $D+m$ dimension, let $\vs$ denote the position vector along the final $m$ dimensions, such that the full position of a site is $(\vr,\vs)$.
Similarly, in the polynomial notation, let $\vy$ denote the variables corresponding to the final $m$ dimensions.
For each $\vs_0$, the fractalized operator $A^{\mathrm{frac}}_{\vs_0}$ is then defined as
\begin{equation}
A^{\mathrm{frac}}_{\vs_0} = \sigma_X \left[\vx^{\vr_0} \vy^{\vs_0} \mathbf{a}(\mathrm{f}(\vy)\circ \vx)\right]\label{eq:afracpoly}
\end{equation}
where
\begin{equation}
\label{eq:fdotx}
\mathrm{f}(\vy)\circ \vx\equiv \left\{f_i(\vy) x_i\right\}_{i\in[1\dots D]}.
\end{equation}
Thus, fractalization maps a $D$-dimensional operator to a set of $D+m$-dimensional operators, $A\rightarrow \{A^{\rmfrac}_{\vs_0}\}$, one for each shift $\vs_0$ in the new dimensions.
Most of the examples in this paper will involve only $m=1$ new dimension.

Similarly, suppose we start out with the $Z$ Pauli operator,
\begin{equation}
B = \sigma_Z\left[\vx^{\vr_0}\mathbf{b}(\bar{\vx})\right],
\end{equation}
where the anchor point $\vr_0$ is chosen similarly as before but spatially inverted: each $(\vr_0)_i$ is minimized subject to the constraint that $\mathbf{b}(\bar{\vx})$ contains only positive powers of $\bar{\vx}$.
Then, the fractalized operator is defined as
\begin{equation}
B^{\mathrm{frac}}_{\vs_0} = \sigma_Z \left[\vx^{\vr_0} \vy^{\vs_0} \mathbf{b}(\mathrm{f}(\bar{\vy})\circ \bar{\vx})\right]
\label{eq:bfracpoly}
\end{equation}

For completeness, let us also express fractalization using the vectoral representation.
The input LCAs are given by matrices $\{\mathbf{F}_i=F_{\vs,\vs^\prime}^{(i)}\}_{i\in[1\dots D]}$.
Starting with an $X$ operator
\begin{equation}
A = \sigma_X\left[a_{\vr,n}\right],
\end{equation}
we find the anchor point $\vr_0$ by maximizing each $(\vr_0)_i$ subject to the constraint that $a_{\vr,n}=0$ for any $(\vr - \vr_0)_i<0$.
Then, the $D+m$ dimensional fractalized operator is 
\begin{equation}
\begin{split}
A^{\rmfrac}_{\vs_0} &= 
\sigma_X\left[a^{\rmfrac}_{(\vr,\vs),n}\right]\\
&=
\sigma_X\left[\left(\prod_{i}\mathbf{F}^{(\vr-\vr_0)_i}_i\right)_{\vs,\vs_0}a_{\vr,n}\right].
\end{split}\label{eq:Afrac}
\end{equation}
where  the ordering of each $\matr{F}_i$ in the product is arbitrary, as all $\matr{F}_i$ commute due to translation invariance (this is obvious from the polynomial representation, $f_i(x)f_j(x)=f_j(x)f_i(x)$).  
Note that since $a_{\vr,n}=0$ if $(\vr-\vr_0)_i<0$,  $\mathbf{F}_i$ is always raised to a non-negative power.

Similarly, for a $Z$ operator,
\begin{equation}
B = \sigma_Z\left[b_{\vr,n}\right],
\end{equation}
we must choose the anchor $\vr_0$ such that each $(\vr_0)_i$ is minimized subject to the constraint that $b_{\vr,n}=0$ for any $(\vr_0-\vr)_i<0$.
Then,
\begin{equation}
B^{\rmfrac}_{\vs_0} = 
\sigma_Z\left[\left(\prod_{i}\mathbf{F}^{(\vr_0-\vr)_i}_i\right)_{\vs_0,\vs}
b_{\vr,n}\right].
\end{equation}

Given a CSS stabilizer group $\mathcal{S}=\langle \{\mathcal{O}_l\}\rangle$, or Hamiltonian
\begin{equation}
H = -\sum_l \mathcal{O}_l,
\end{equation}
where $\mathcal{O}_l$ are local $X$ or $Z$ generators of the stabilizer group,
we may define the fractalized stabilizer group $\mathcal{S}^\rmfrac=\langle \{\mathcal{O}^\rmfrac_{\vs,l}\}\rangle$, or Hamiltonian
\begin{equation}
H^{\rmfrac} = -\sum_{\vs,l} \mathcal{O}^{\rmfrac}_{\vs,l} 
\end{equation}
which, as we will show, inherits many of the properties of $H$.
Note that $H^{\rmfrac}$ is always translation invariant along the final $m$ dimensions by construction.

Finally, fractalization may also be applied to non-local operators, provided certain conditions are satisfied.
If the original system has open boundary conditions, fractalization can be applied straightforwardly.
In the case of periodic boundary conditions, however, it may not be possible to choose an anchor point consistently for a non-local operator.
Suppose an operator is non-local in the $i$th direction.
If $f_i(y)^{L_i}=1$, then the fractal is commensurate with the system size and the anchor $r_{0,i}$ can be chosen arbitrarily.
If this is not the case, then we must find a set of polynomials $Q_i=\{ q(\vy) : q(\vy)f_i(\vy)^{L_i} = q(\vy)\}$, and make a choice of basis polynomials $\{q_j(\vy)\}_j$ for $Q_i$.    
Then, an operator $A$ or $B$ maps on to a set of fractalized operators, one for each $q_j(\vy)$.
Eq~\ref{eq:afracpoly} and \ref{eq:bfracpoly} generalize to
\begin{equation}
\begin{split}
A^{\rmfrac}_j &= \sigma_X\left[\vx^{\vr_0} q_j(\vy) \mathbf{a}(\vf(\vy)\circ\vx)\right]\\
B^{\rmfrac}_j &= \sigma_Z \left[\vx^{\vr_0} q_j(\bar{\vy}) \mathbf{b}(\mathrm{f}(\bar{\vy})\circ \bar{\vx})\right].
\end{split}
\end{equation}
where the choice of $r_{0,i}$ can be arbitrary.
Thus, each operator only maps on to $\log_2|Q_i|$ fractalized operators.
If an operator is non-local in multiple directions $i\in I$, then $q_j(\vy)$ must be chosen from $\cap_{i\in I} Q_i$ such that the fractal is commensurate with all directions.

\subsection{Properties}
\subsubsection{Locality}
We first point out that fractalization is locality preserving.  
As long as $A$ has support only on a local patch of the $d$-dimensional lattice, $A^{\rmfrac}_{\vs_0}$ will also have support on a local patch of the $D+m$-dimensional lattice near $\vs_0$.
This follows from the locality of the LCA rules: $F^{(i)}_{\vs,\vs_0}$ is only non-zero for small $|\vs-\vs_0|$.

Non-local operators map on to non-local, potentially fractal, operators.
To see what this means, consider the $D=1$ chain with length $L$ and open boundary conditions, and the $1$-dimensional LCA rule $f(y)=1+y$.
The non-local operator 
\begin{equation}
S =  \prod_{r} X_r =\sigma_X\left[\sum_{r} x^r\right] 
\end{equation}
maps on to the fractal operators
\begin{equation}
S^{\rmfrac}_{s} = \sigma_X\left[y^s \sum_{r} f(y)^r x^r\right].
\end{equation}
in two dimensions.
Recall that $f(y)^n$ generates the $n$th row of the Sierpinski triangle fractal (Eq~\ref{eq:sierpinski}).
Hence, $S^{\mathrm{frac}}_s$ is an operator that acts as $X$ on sites along a Sierpinski triangular fractal subsystem, and $s$ just denotes an overall shift in the $y$ direction.
This operator has support on a fraction of sites scaling as $L^{d_f}$, where $d_f\approx 1.58$ is the Hausdorff dimension of the Sierpinski triangle.

\subsubsection{Commutativity}
The next property of fractalization is commutativity preservation.  
Suppose we have two operators ${A=\sigma_X[\vx^{\vr_0}\mathbf{a}(\vx)]}$ and ${B=\sigma_Z[\vx^{\vr_1}\mathbf{b}(\bar{\vx})]}$ satisfying $[A,B]=0$.
Then, $[A^{\rmfrac}_{\vs_0},B^\rmfrac_{\vs_1}]=0$ for all $\vs_0,\vs_1$ as well.
This can be seen by computing the commutation polynomial.
The commutation polynomial for $A$ and $B$ is 
\begin{equation}
c(\vx) = \vx^{\vr_0-\vr_1}\mathbf{a}^T(\vx) \mathbf{b}(\vx),
\end{equation}
which has a zero constant term iff $[A,B]=0$.
The commutation polynomial for $A^{\rmfrac}_{\vs_0}$ and $B^{\rmfrac}_{\vs_1}$ is
\begin{equation}
\begin{split}
c^{\rmfrac}(\vx,\vy) &= \vx^{\vr_0-\vr_1}\vy^{\vs_0-\vs_1}\mathbf{a}^T(\vf(\vy)\circ\vx)\mathbf{b}(\vf(\vy)\circ \vx)\\
&=\vy^{\vs_0-\vs_1} c(\vf(\vy)\circ \vx)
\end{split}
\end{equation}
Therefore, if $c(x)$ has a zero constant term, $c^\rmfrac(\vx,\vy)$ also has a zero constant term regardless of $\vs_0,\vs_1$.

It is also instructive to work in the vectoral representation.  
Let $AB = (-1)^c B A$, 
then, commutativity of $A=\sigma_X[(a_{\vr,n})]$ and $B=\sigma_Z[(b_{\vr,n})]$ implies that
\begin{equation}
c = \sum_{\vr,n}  b_{\vr,n}a_{\vr,n} 
\end{equation}
is zero.
The fractalized operators satisfy $[[A^\rmfrac_{\vs_0},B^\rmfrac_{\vs_1}]] =(-1)^{c_{\vs_0,\vs_1}^\rmfrac} $
where
\begin{equation}
c^\rmfrac_{\vs_0,\vs_1} =  \left(\prod_{i} \mathbf{F}_i^{(\vr_1-\vr_0)_i}\right)_{\vs_1,\vs_0} c
\label{eq:fraccomm}
\end{equation}
which is zero if $c=0$.  
If instead $A$ and $B$ anticommute, then $c=1$ and $c^\rmfrac_{\vs_0,\vs_1}$ may be zero or non-zero depending on their separation $\vs_1-\vs_0$.

\subsection{Three-step process}\label{sec:interp}
We now present a three-step process which is equivalent to fractalization, but offers insight into the relation between the original and fractalized operators.
The three steps are $1)$ constructing a layered system, $2)$ unitary transformation, and $3)$ choosing a different set of generators.
Each of these three steps is explained in detail below.
While fractalization may be applied to systems of arbitrary size and LCA rules, the three-step process is only applicable under certain conditions:
the boundary conditions of the $D$ original directions must be either open or periodic with $\matr{F}_i^{L_i}=\mathbb{1}$,
 $\matr{F}_i$ must all be invertible, and
 the boundary conditions along the $m$ new directions must be periodic.

Let us start with an $X$ or $Z$ operator in the vectoral representation, $A=\sigma_X[a_{\vr,n}]$ or $B=\sigma_Z[b_{\vr,n}]$, defined in $D$ dimensions, and choose an anchor point $\vr_0$ in the same way as before.
The first step is to extend these operators to a layered system, with each layer labeled by its coordinate $\vs$ along the perpendicular direction.  
The operator $A$ or $B$ now maps on to a set of operators on each layer separately,
\begin{equation}
A_{\vs_0} = \sigma_X[a_{\vr,n}\delta_{\vs,\vs_0}],\;\;\;\;
B_{\vs_0} = \sigma_Z[b_{\vr,n}\delta_{\vs,\vs_0}].
\end{equation}
on the $D+m$ dimensional lattice.

The next step involves a non-local (along the $\vs$ directions) unitary transformation.  
The unitary transformation involves a series of controlled-$X$ (CX) gates.  
In particular, for any invertible binary matrix $\mathbf{M}$ with $M_{ii}=1$, there is a corresponding CX circuit which implements the transformation
\begin{equation}
\sigma_X[\mathbf{v}]\rightarrow \sigma_X[\mathbf{M}\mathbf{v}], \;\;\;\;
\sigma_Z[\mathbf{w}]\rightarrow \sigma_Z[\mathbf{M}^{-1,T}\mathbf{w}].
\end{equation}
In Appendix~\ref{app:cx}, we give a general algorithm for determining a CX circuit for any such $\matr{M}$.  
Here, we choose the matrix 
\begin{equation}
M_{(\vr,\vs,n),(\vr^\prime,\vs^\prime,n^\prime)} = \delta_{\vr,\vr^\prime}\delta_{n,n^\prime} \left(\prod_i \mathbf{F}_i^{r_i}\right)_{\vs,\vs^\prime} 
\end{equation}
which maps $A_{\vs_0}\rightarrow\tilde{A}_{\vs_0},~B_{\vs_0}\rightarrow\tilde{B}_{\vs_0}$,
\begin{equation}
\begin{split}
\tilde{A}_{\vs_0} = \sigma_X\left[\left(\prod_{i} \mathbf{F}_i^{r_i}\right)_{\vs,\vs_0} a_{\vr,n}\right],\\
\tilde{B}_{\vs_0} = \sigma_Z\left[\left(\prod_{i} \mathbf{F}_i^{-r_i}\right)_{\vs_0,\vs} b_{\vr,n}\right].
\end{split}
\end{equation}
which are highly non-local operators along the $\vs$ directions as they involve $\mathbf{F}_i$ being raised to potentially high powers $r_i$.

The third and final step involves choosing a different set of generators for the groups $\langle \{\tilde{A}_{\vs_0}\}\rangle$ and $\langle \{\tilde{B}_{\vs_0}\}\rangle$.  
For $X$ operators we choose
\begin{eqnarray}
A^{\rmfrac}_{\vs_0} &=& \prod_{\vs^\prime} \tilde{A}_{\vs^\prime}^{\left(\prod_i \mathbf{F}^{-r_{0,i}}_i\right)_{\vs^\prime,\vs_0}}
\label{eq:genchange}
\\
&=& \sigma_X \left[\sum_{\vs^\prime} \left(\prod_i \matr{F}_i^{-r_{0,i}}\right)_{\vs^\prime,\vs_0} \left(\prod_i\matr{F}_i^{r_i}\right)_{\vs,\vs^\prime} a_{\vr,n}\right]\nonumber\\
&=& \sigma_X \left[\left(\prod_i \matr{F}_i^{(\vr-\vr_0)_i}\right)_{\vs,\vs_0}  a_{\vr,n}\right]
\end{eqnarray}
which matches Eq~\ref{eq:Afrac}.
Similarly, for $Z$ operators,
\begin{equation}
\begin{split}
B^{\rmfrac}_{\vs_0} &= \prod_{\vs^\prime} \tilde{B}_{\vs^\prime}^{\left(\prod_i \mathbf{F}^{r_{0,i}}_i\right)_{\vs_0,\vs^\prime}}
\end{split}
\end{equation}
results in the fractalized operator.
Since $\matr{F}_i$ are all invertible, this just corresponds to a different choice of generators for their generated groups.  
Although $\tilde{A}_{\vs_0}$ were highly non-local, $A^\rmfrac_{\vs_0}$ are all local operators.
Thus, locality is restored.

This three-step process has important implications for fractalized quantum codes.  
Consider the CSS stabilizer Hamiltonian 
\begin{equation}
H = -\sum_l \mathcal{O}_l
\end{equation}
with the $2^k$-dimensional ground state manifold obtained as the simultaneous $+1$ eigenstate of all $\mathcal{O}_l$.  
Let us choose LCA rules and the system sizes such that the three-step process can be applied.  
After the first step, $H\rightarrow \sum_{\vs} H_{\vs}$ simply describes a layered system with ground state degeneracy $2^{L k}$. 
The second step is a unitary transformation, $H\rightarrow U H U^\dagger$, which does not affect the ground state degeneracy.  
The ground states themselves are related to the original ground states of the layered system by the non-local unitary transformation $\ket{\psi}\rightarrow U\ket{\psi}$. 
Finally, since all ground states are $+1$ eigenstates of each term, choosing a different set of generators for the stabilizer group does not affect the ground state manifold (although higher excited eigenstates are shifted).
Thus, starting with a locally generated CSS stabilizer Hamiltonian with ground state degeneracy $2^k$, the fractalized Hamiltonian will have a $2^{L k}$ degenerate ground state manifold which is unitarily related to those of the layered system by the non-local unitary transformation $U$. 
In particular, this implies that upon compactifying the additional $m$ dimensions the fractalized model is in the same $D$ dimensional gapped quantum phase of matter as the layered system.

\subsection{Generalizations}
\subsubsection{Beyond qubits}
Everything discussed above straightforwardly generalizes from qubit degrees of freedom to $\mathbb{Z}_p$ qudits of prime dimension $p$.  
The $X$ and $Z$ Pauli operators are replaced by $\mathbb{Z}_p$ clock operators, satisfying the relation 
$XZ = e^{\frac{2\pi i}{p}} ZX$.  

Fractalization now takes as input $p$-state LCA rules, specified by polynomials $f(\vx)$ with $\mathbb{F}_p$ coefficients.  
Such LCA still generate fractal structures due to the property $f(\vx)^{p^n}=f(\vx^{p^n})$, where self-similarity occurs at scales $p^n$.
Our previous discussions readily applies, with arithmetic now done modulo $p$.

\subsubsection{Higher order LCA}\label{sec:higherorder}
The fractalization procedure generalizes to higher order LCA.  
In an $n$th order LCA, the state at time $t$, $c_t$, may be determined by the states at times back to $t-n$,
\begin{equation}
c_t(\vy) = \sum_{j=1\dots n} f^{(j)}(\vy) c_{t-j} (\vy)
\end{equation}
where the LCA update rule is now specified by a list of $n$ polynomials $\{f^{(j)}(\vy)\}$.

Fractalization from $D$ to $D+m$ dimensions now takes as input $D$, $m$-dimensional, $n$-th order LCA rules, $\{f^{(j)}_i(\vy)\}$.  
One possible generalization for  
Eq.~\ref{eq:Afrac} is
\begin{equation}
\begin{split}
A^{\rmfrac}_{\vs_0} =& \sigma_X\left[
\vx^{\vr_0} \vy^{\vs_0} \mathbf{a}\left(\left\{\sum_{j} {f}^{(j)}_i(\vy)x_i^j\right\}_{i}\right)
\right] 
\end{split}\label{eq:hoAfrac}
\end{equation}
and 
\begin{equation}
\begin{split}
B^{\rmfrac}_{\vs_0} =& \sigma_Z\left[
\vx^{\vr_0} \vy^{\vs_0} \mathbf{b}\left(\left\{\sum_{j} {f}^{(j)}_i(\bar{\vy})\bar{x}_i^j\right\}_i\right)
\right] 
\end{split}
\end{equation}
This generalization is chosen such that the fractalized Ising model (see Sec~\ref{sec:ising}) has as its ground state the set of valid space-time LCA trajectories.
However, this higher order generalization no longer has as many of the nice properties as before. 

For one, $[A,B]=0$ is no longer a sufficient criterion for the commutativity of the fractalized operators.
Instead, commutativity is only guaranteed if all translations commute.
Let $A_{\vr}$ and $B_{\vr}$ be translations of $A$ and $B$, such that their anchor point is at $\vr$.
Then, $[A_{\vr_0},B_{\vr_1}]=0$ for all $\vr_{0/1}$, implies that $[A^{\rmfrac}_{\vr_0,\vs_0},B^{\rmfrac}_{\vr_1,\vs_1}]=0$ for all $\vr_{0/1},\vs_{0/1}$.
This follows from the fact that, for operators whose translations all commute, 
\begin{equation}
\mathbf{a}^T(\vx)\mathbf{b}(\vx)=\mathbf{a}^T(\vf(\vy)\circ \vx)\mathbf{b}(\vf(\vy)\circ \vx)=0 \, .
\end{equation}
Thus, fractalization with higher order LCA should only be applied for translation invariant commuting codes.

We also do not have a generalization of the three-step process for fractalizing higher order LCA, even for translation invariant codes.
Thus, the precise connection between the original and fractalized code for a general higher order LCA remains an open question.

\section{Examples and Discussion}\label{sec:examples}

In this section, we go through a number of examples of fractalization applied to well-understood models, connecting them with more exotic fractal models in the literature.
We start with the 1D quantum Ising model, which we show fractalizes to a spin model studied by Newman and Moore~\cite{newmanmoore} as a translation invariant model of glassiness, and also as a classical code~\cite{CAcode,yoshidaclassical,Nixon2020}.
We then move on to the cluster model~\cite{cluster} and find that it fractalizes into cluster states on higher dimensional lattices, known examples of fractal subsystem SPT phases~\cite{fractalSSPT} which have been shown to be useful for universal measurement-based quantum computation~\cite{Raussendorf_2019,devakul2018universal}.
We then fractalize Kitaev's 2D toric code model~\cite{kitaev}, which results in Yoshida's fractal spin liquids~\cite{yoshida2013exotic}, a family of models of type-II fracton topological order.

\subsection{1D Ising model}\label{sec:ising}
The 1D Ising model is defined on a chain with one qubit per site $r$.  
The  Hamiltonian simply consists of pairwise nearest neighbor Ising interactions 
\begin{equation}
H_\mathrm{1DIsing} = -\sum_{r} Z_{r-1} Z_r = -\sum_r \sigma_Z[x^r (1+\bar{x})]
\end{equation}
which, after fractalization with a $1$ dimensional LCA $f(y)$, becomes
\begin{equation}
H_\mathrm{1DIsing}^\rmfrac = -\sum_{r,s} \sigma_Z[x^ry^s (1+f(\bar{y})\bar{x})].
\end{equation}
Using the choice $f(y)=1+y$, we arrive at the Newman-Moore model~\cite{newmanmoore},
\begin{equation}
H_\mathrm{NM} = -\sum_{r,s} Z_{(r,s)} Z_{(r-1,s)} Z_{(r-1,s-1)}.
\end{equation}
where $(r,s)$ labels the $x,y$ coordinates on the square lattice.
The global $\mathbb{Z}_2$ symmetry of the Ising model maps on to a set of fractal symmetries,
\begin{equation}
S=\sigma_X\left[\sum_r x^r\right] \rightarrow S^{\rmfrac}_{s} =\sigma_X\left[y^s \sum_r f(y)^r x^r\right]
\end{equation}
which act on a subsystem of sites corresponding to the Sierpinski triangle fractal, shifted by $s$ along the $y$ direction.

The ground states of $H_{\mathrm{1DIsing}}^\rmfrac$ are spin configurations that are valid space-time LCA trajectories.
Let us denote a product state in the $Z_{(r,s)}$ basis by $|\{z_{(r,s)}\}\rangle$, where  $Z_\vr|\{z_{(r,s)}\}\rangle=(-1)^{z_{(r,s)}}$.
Then, $z_{(r,s)}$ forms a valid LCA trajectory, with $r$ playing the role of time, and $s$ playing the role of space.
This remains true in higher dimensions, as well as for higher-order LCA.  
For the $D$-dimensional Ising model, fractalized with $m$-dimensional LCA, the resulting $D+m$-dimensional $H_{\mathrm{1DIsing}}^\rmfrac$ has as its ground state valid space-time LCA trajectories,
where $\vr$ are time dimensions, and $\vs$ are space dimensions.  
Multiple time dimensions, each governed by their own update rule, are possible since translation-invariant LCA rules commute.

The Ising chain may also be viewed as the classical repetition code, whereby the logical bit is represented by the two symmetry-broken ground states.
Similarly, the fractalized Ising chain, in $1+m$ dimensions, is also useful as a classical code~\cite{yoshidaclassical,CAcode,Nixon2020}, in which logical bits are again represented by the symmetry-broken ground states, of which there are now $2^{\mathcal{O}(L^m)}$.
The fractal symmetry operators $S^{\rmfrac}_\vs$ are logical operators that flip between the  symmetry-broken ground states.
Upon generalizing to $\mathbb{Z}_p$ degrees of freedom, this code saturates an information storage tradeoff bound~\cite{BravyiPoulinTerhal} in the limit $p\rightarrow\infty$~\cite{yoshidaclassical}.

The nature of the excitations of the fractalized Ising model can be deduced from the excitations of the Ising model.
The Ising model in 1D has point-like domain-wall excitations, which are created from a ground state $\ket{\psi}$ by flipping a segment of spins, $W\ket{\psi}\equiv \prod_{r\in[\ell_1\dots \ell_2-1]}X_r \ket{\psi}$.
Let $B_r=Z_{r-1}Z_r$ be a term in $H_{\mathrm{1DIsing}}$, which satisfies the commutation relation
\begin{equation}
[[W,B_r]] = (-1)^{\delta_{r,\ell_1} + \delta_{r,\ell_2}} \, ,
\end{equation}
making it clear that $W$ only creates two excitations at $\ell_1$ and $\ell_2$ when acting on the ground state.  
From Eq.~\ref{eq:fraccomm}, the fractalized versions of these operators satisfy
$[[W^\rmfrac,B^{\rmfrac}_{r,s}]] = (-1)^c$ with
\begin{equation}
c=\delta_{r,\ell_1}\delta_{s,s_1} + \delta_{r,\ell_2}(\matr{F}^{\ell_2-\ell_1})_{s,s_1} \, .
\end{equation}
Therefore, $B^{\rmfrac}_{s_1}$ creates a single excitation at $(\ell_1,s_1)$, 
while creating potentially many excitations along $r=\ell_2$, depending on $\matr{F}^{\ell_2-\ell_1}$.
However, if $\ell_2-\ell_1=2^n$, ${f(y)^{2^n} = f(y^{2^n})}$ and so $(\matr{F}^{2^n})_{s_1,s}$ is only non-zero for a few $s$; for $f(y)=1+y$, only two further excitations are created at $(\ell_1+2^n,s_1)$ and $(\ell_1+2^n,s_1+2^n)$.  
The excitations of $H^{\rmfrac}_{\mathrm{1DIsing}}$ are therefore point-like, and may be created in special configurations separated by distances $2^n$ via fractal operators.  
They are an example of (non-topological) fractons.  

This can be generalized to higher dimensions.  
The 2D Ising model, for example, has loop-like domain wall excitations and is treated in section~\ref{sec:fractalizingloops}.  

\subsection{1D Cluster model}\label{sec:clusterstate}
The 1D cluster model, in CSS form, has two qubits per site and is described by the Hamiltonian
\begin{equation}
\begin{split}
H_{\mathrm{clus}} &= -\sum_{r} X_r^{(1)} X_r^{(2)} X_{r+1}^{(1)} - \sum_r Z_{r-1}^{(2)} Z_r^{(1)} Z_r^{(2)}\\
&= -\sum_{r}\sigma_X\left[x^r\begin{pmatrix}1+x\\1\end{pmatrix}\right]
-\sum_{r}\sigma_Z\left[x^r\begin{pmatrix}1\\1+\bar{x}\end{pmatrix}\right] .
\end{split}
\end{equation}
After fractalization with a $1$-dimensional LCA $f(y)$,
\begin{equation}
\begin{split}
H_{\mathrm{clus}}^\rmfrac = 
& -\sum_{r,s}\sigma_X\left[x^r y^s\begin{pmatrix}1+f(y)x\\1\end{pmatrix}\right]\\
&-\sum_{r,s}\sigma_Z\left[x^r y^s\begin{pmatrix}1\\1+f(\bar{y})\bar{x}\end{pmatrix}\right] ,
\end{split}
\end{equation}
which describes another cluster state on a 2D bipartite lattice.
With the choice $f(y)=1+y$, for example, $H^{\rmfrac}_{\mathrm{clus}}$ describes the cluster state on the honeycomb lattice~\cite{fractalSSPT}.
Higher dimensional models may be created by using $m$-dimensional LCA rules,
for example, using the $2$-dimensional LCA rule $f(\vy)=1+y_1+y_2$, one arrives at a 3-dimensional cluster state, with Sierpinski tetrahedron fractal subsystem symmetries.

The cluster model describes a non-trivial symmetry-protected topological (SPT) phase, protected by the ${\mathbb{Z}_2\times\mathbb{Z}_2}$ symmetry that is generated by $S_1=\prod_r Z_r^{(1)}$ and $S_2=\prod_r X_r^{(2)}$.
The fractalized cluster model also describes an SPT phase, protected by the set of fractal subsystem symmetries generated by $S^{\rmfrac}_{\alpha,s}$, $\alpha\in\{1,2\}$.

The non-triviality of $H^{\rmfrac}_{\mathrm{clus}}$ as a fractal subsystem SPT follows from that of $H_{\mathrm{clus}}$.  
The cluster chain $H_{\mathrm{clus}}$ realizes a non-trivial SPT phase, as can be observed from the fact that, at a boundary, the $G=\mathbb{Z}_2\times\mathbb{Z}_2$ symmetry acts as a non-trivial projective representation~\cite{PhysRevB.84.235128}.
We may apply fractalization directly to the open cluster chain and its symmetry operators, following the three-step process.
In the first step, the system is composed of decoupled cluster chains along $x$, each with their own symmetry $S_{\alpha,s}$  realizing a projective representation on the left edge.  
The second step is a unitary transformation, local along $x$, which does not affect the projective representation.  
The third step is a different choice of generators for the symmetry group $\langle \{ S_{\alpha,s}\}\rangle$. 
We can simply choose the origin such that the anchor point for $S_{\alpha,s}$ is $r_0=0$, in which case the change of generators (Eq~\ref{eq:genchange}) is trivial.  
Thus, $S^{\rmfrac}_{\alpha,s}$ realizes the same projective representation in $H^{\rmfrac}_{\mathrm{clus}}$ as $S_{\alpha,s}$ from the layered $H_{\mathrm{clus}}$.  

If we take periodic boundary conditions along the $x$ direction as well, along with an irreversible LCA, then the three-step process cannot be applied.
Following the discussion in Sec.~\ref{sec:fracdef}, the symmetry group of $H^{\rmfrac}_{\mathrm{clus}}$ is generated by $\{S_{\alpha,j}^{\rmfrac}\}$, which is smaller than the symmetry group of the layered system (or may even be trivial in the most extreme case), depending on the number of solutions to $q(y)f(y)=q(y)$.
These are pseudo-SPT phases, as defined in Ref.~\onlinecite{Devakul2018}.

For a given set of fractal symmetries, there are an infinite number of non-trivial SPT phases~\cite{Devakul2018}.
The fractal subsystem SPT phase obtained from fractalizing the cluster state is only the simplest non-trivial phase.
The remaining phases are generically not strictly translation invariant (they are translation invariant with a larger period $2^n$), and therefore cannot be realized by fractalization.

\subsection{2D toric code model}\label{sec:toriccode}
The 2D toric code~\cite{kitaev} is defined on the square lattice with a qubit on every bond.
Associating with each site the two qubits on bonds straddling it in the $+\mathbf{x}_1$ and $+\mathbf{x}_2$ direction,  the Hamiltonian may be written as
\begin{align}
H_{\mathrm{TC}} =&
-\sum_{\vr} \sigma_X\left[\vx^\vr\begin{pmatrix}1+x_2\\1+x_1\end{pmatrix}\right] 
-\sum_{\vr} \sigma_Z\left[\vx^\vr\begin{pmatrix}1+\bar{x}_1\\1+\bar{x}_2\end{pmatrix}\right]
\nonumber \\
=& -\sum_\vr A_\vr - \sum_\vr B_\vr \, .
\end{align}
Applying fractalization with two 1-dimensional LCA rules, $f_1(y),f_2(y)$, we get
\begin{equation}
\begin{split}
H^{\rmfrac}_{\mathrm{TC}} = &
-\sum_{\vr,s} \sigma_X\left[\vx^\vr y^s\begin{pmatrix}1+f_2(y)x_2\\1+f_1(y)x_1\end{pmatrix}\right] \\
&-\sum_{\vr,s} \sigma_Z\left[\vx^\vr y^s \begin{pmatrix}1+f_1(\bar{y})\bar{x}_1\\1+f_2(\bar{y})\bar{x}_2\end{pmatrix}\right].
\end{split}
\label{eq:fractc}
\end{equation}
which is Yoshida's fractal spin liquid model~\cite{yoshida2013exotic}.
We take $H_{\mathrm{TC}}$ on an $L_1\times L_2$ torus, and $H^{\rmfrac}_{\mathrm{TC}}$ on an ${L_1\times L_2\times L_3}$ $3$-torus.

The logical operators of the toric code, 
\begin{equation}
\begin{split}
\ell^X_{1} = \sigma_X\left[\sum_{r_1}\begin{pmatrix} x_1^{r_1} \\ 0\end{pmatrix}\right]\;\;\;
\ell^Z_{1} = \sigma_Z\left[\sum_{r_2}\begin{pmatrix} \bar{x}_2^{r_2} \\ 0\end{pmatrix}\right]\\
\ell^X_{2} = \sigma_X\left[\sum_{r_2}\begin{pmatrix} 0\\x_2^{r_2}\end{pmatrix}\right]\;\;\;
\ell^Z_{2} = \sigma_Z\left[\sum_{r_1}\begin{pmatrix} 0 \\ \bar{x}_1^{r_1}\end{pmatrix}\right]
\end{split}
\end{equation}
generate the $2$ qubit Pauli algebra on the ground state manifold of $H_{\mathrm{TC}}$.
If $f_1(y)^{L_1}=f_2(y)^{L_2}=1$, then they can be straightforwardly fractalized,
\begin{equation}
\begin{split}
\ell^{X,\rmfrac}_{1,s} = \sigma_X\left[y^s\sum_{r_1}\begin{pmatrix} f_1(y)^{r_1}x_1^{r_1} \\ 0\end{pmatrix}\right]\\
\ell^{Z,\rmfrac}_{1,s} = \sigma_Z\left[y^s\sum_{r_2}\begin{pmatrix} f_2(\bar{y})^{r_2}\bar{x}_2^{r_2} \\ 0\end{pmatrix}\right]
\end{split}
\end{equation}
and similarly for $\ell^{X/Z,\rmfrac}_{2,s}$.
It can be easily checked that $[[\ell^{X,\rmfrac}_{\alpha,s_0},\ell^{Z,\rmfrac}_{\beta,s_1}]]=(-1)^{\delta_{\alpha \beta}\delta_{s_0 s_1}}$,
 thus generating a $2 L_3$ qubit Pauli algebra on the ground state manifold.
 If $f_i(y)^{L_i}\neq 1$, then some smaller subgroup is realized instead.

The toric code describes a 2D topologically ordered phase.  
The fractalized toric codes describe a family of (type-II) fracton topological ordered phases.
Fracton topological order is characterized by a subextensive ground state degeneracy (scaling with an envelope $2^{\mathcal{O}(L)}$) and quasiparticle excitations with restricted mobility.  
When the three-step process for fractalization holds, the ground state degeneracy of $H^{\rmfrac}_{\mathrm{TC}}$ is $2^{2 L_3}$, but will be less for generic system sizes where $f_i(y)^{L_i}\neq 1$.  

The excitations of the toric code are created at the ends of string operators.  
Fractalizing these string operators and following a similar analysis as in Sec.~\ref{sec:ising}, one finds that the excitations of the fractalized toric code are also point-like and must be created in particular configurations.
As pointed out in Ref.~\onlinecite{yoshida2013exotic}, these quasiparticle excitations are only strictly immobile  if $f_1(y)$ and $f_2(y)$ are algebraically unrelated.
Two polynomials are algebraically related if there is a non-trivial solution to
\begin{equation}
f_1(y)^{n_1} =  c f_2(y)^{n_2}
\label{eq:algebraicrelation}
\end{equation}
for any finite constants $n_i$,$c$, without periodic boundary conditions.

Haah's cubic code~\cite{Haah} can also be viewed as a higher order fractal spin liquid~\cite{yoshida2013exotic}. 
Using the generalization of fractalization to higher order LCA, with 
\begin{equation}
\begin{split}
f_1^{(1)}(y)&=1+y+y^2\\
f_1^{(2)}(y)&=0\\
f_2^{(1)}(y)&=1+y\\
f_2^{(2)}(y)&=1+y+y^2
\end{split}
\end{equation}
one arrives at a model that is locality preserving unitarily related~\cite{yoshida2013exotic} to Haah's cubic code, up to a redefinition of the lattice vectors.

The toric code may also be regarded as a limit of the Ising gauge theory (IGT), which is obtained by gauging the global $\mathbb{Z}_2$ symmetry of the 2D quantum Ising model $H_{\mathrm{2D Ising}}$.  
After gauge fixing, the IGT takes the form of a perturbed toric code,
\begin{equation}
H_{\mathrm{IGT}} =-h\sum_\vr A_\vr -K\sum_\vr B_\vr - J\sum_{\vr,n}Z_\vr^{(n)} - \Gamma\sum_{\vr,n}X_{\vr}^{(n)}
\end{equation}
where $n=1,2$ labels the qubits on each site.
$H_{\mathrm{IGT}}$ is in the topologically ordered phase when $J/h, \Gamma/K \ll 1$.
The fractalized IGT, $H^{\rmfrac}_{\mathrm{IGT}}$, can either be obtained by fractalizing $H_{\mathrm{IGT}}$ or alternatively by applying a generalized gauging procedure~\cite{williamsongauge,vijay2016fracton,shirley2018FoliatedFracton} to the fractal subsystem symmetries of the fractalized 2D Ising model.
Thus, the generalized gauging procedure naturally arises in this context as the ``fractalized'' version of the usual gauging procedure.

In Refs.~\onlinecite{Dua2019,Williamson2018b}, the effect of dimensional compactification of 3D fracton models was investigated. 
Compactifying a 3D translation invariant type-II fracton code along one dimension leads to a 2D code which is equivalent to copies of toric code and trivial states by a locality-preserving unitary transformation~\cite{haah2016algebraic,Haah2018a,bombin2014structure}.
Take the $L_1\times L_2\times L_3$ fractalized toric code models, and consider compactifying along the third dimension.
In cases where the three-step process for fractalization can be applied, we can easily show that the resulting compactified model is local unitarily equivalent to $L_3$ copies of the toric code:
The first step creates the layered toric codes, the second step is a unitary transformation, and the third step is an irrelevant change of generators for the stabilizer group.
The unitary transformation, the most important step, is non-local only in the third direction.
Thus, once the third dimension is compactified, the second step is a local unitary transformation which transforms the ground state into $L_3$ copies of the toric code ground state.

In a recent manuscript~\cite{coupledlayers} by the present authors, a coupled cluster state construction was presented for a number of models, including the toric code and many type-II fracton phases.
The Hamiltonian takes the form $H=H_{\mathrm{clus}} + \lambda H_{\mathrm{coup}}$, where $H_{\mathrm{clus}}$ is the cluster Hamiltonian on decoupled chains or layers, and $H_{\mathrm{coup}}$ couples them together.  
By choosing the the cluster states and couplings appropriately, as discussed in Ref.~\onlinecite{coupledlayers}, 
the low energy effective Hamiltonian in the limit $\lambda\rightarrow\infty$ coincides with the desired fracton or toric code model.
In particular, a construction of the toric code in terms of coupled wires, and Yoshida's type-II fracton phases~\cite{yoshida2013exotic} in terms of coupled layers was obtained.
This coupled layer construction of Yoshida's models follows directly from fractalization applied to the coupled wire construction of the toric code.
Specifically, the Hamiltonian $H$ describing the coupled wire construction of the toric code can be written in terms of only $X$ or $Z$ operators.
Applying fractalization, $H\rightarrow H^\rmfrac$, the cluster state wires map on to fractal cluster state layers, and in the limit $\lambda\rightarrow\infty$, $H^\rmfrac$ realizes the fractalized toric code, which is exactly Yoshida's models.

\section{Fractalized loop excitations} 
\label{sec:fractalizingloops}

In this section we describe several examples that involve fractalizing stabilizer models with loop-like excitations. 
We find that the excitations in the fractalized models are no longer generically loop or string-like, but form more general extended ``fractalized loop'' excitations.
We determine the criterion for such models to lack any string-like excitations whatsoever.  
This implies the lack of any membrane logical operators and can lead to a boost in the energy barrier for applying a logical operator from $\mathcal{O}(L)$ in the original model to $\mathcal{O}(L^\alpha)$, for some $1<\alpha\leq 2$ determined by the choice of fractalization, where $L$ is the linear size of the system. 

For this section, we make use of a more compact notation for the sum of $K$ Hamiltonian terms,
\begin{equation}
\sum_{k=1\dots K}\sigma_X[\mathbf{a}_k(\vx)] \equiv \sigma_X[\mathbf{A}(\vx)]
\end{equation}
where 
\begin{equation}
\mathbf{A}(\vx) = \begin{pmatrix} \mathbf{a}_1(\vx) & \dots & \mathbf{a}_K(\vx)\end{pmatrix}
\end{equation}
is an $N\times K$ matrix whose columns correspond to each of the Hamiltonian terms (and rows correspond to the physical qubits on each site as before). 
A similar notation is also used for $\sigma_Z[\mathbf{B}(\vx)]$.
This representation is useful for translation invariant models with multiple types of $X$ or $Z$ terms, such as the higher dimensional toric code models.

\subsection{2D quantum Ising model} \label{sec:fracloop2d}

As a warmup we fractalize the 2D quantum Ising model which supports looplike domain wall excitations. 
The Hamiltonian is given by
\begin{align}
H_{\text{2DIsing}} = - \sum_{\textbf{r}} \sigma_Z[\vx^{\vr} \mathbf{B}(\bar{\vx})] 
\end{align}
with
\begin{equation}
\mathbf{B}(\bar{\vx})
= \begin{pmatrix}
	1+\bar{x}_1 & 1+\bar{x}_2 
	\end{pmatrix}
  \, ,
 \label{eq:2DQIBr}
\end{equation} 
where, as explained above, the two columns correspond to the Ising interactions along the $x_1$ and $x_2$ directions.

The logical operators for the code on periodic boundary conditions are given by 
\begin{align}
\label{eq:2DQIL}
\ell^X = 
\sigma_X \left[ 
 \sum_{\textbf{r}}
	\vx^{\vr}  
 \right] \, ,
 &&
\ell^Z = 
 \sigma_Z \left[ 1
 \right] \, ,
 \end{align}
 which represent a global $X$ spin flip and single site $Z$ operator, respectively. 
 Hence, the Ising model is not topologically ordered as the ground space degeneracy can be split locally. 
 
 A rectangular loop excitation can be created with a truncated membrane operator,
 \begin{equation}
 W = \sigma_X\left[\sum_{\vr \in \mathcal{R}} \vx^\vr\right] \equiv \sigma_X[\mathbf{w}(\vx)]\, ,
 \end{equation} 
 where $\mathcal{R}=\mathcal{R}_1\times\mathcal{R}_2$ is a set of sites in some rectangular region $\mathcal{R}_i = \{0\dots l_i-1\}$.
 The excitations created in response to $W$ acting on the ground state is described by the polynomial vector
 \begin{equation}
 \mathbf{E}_{W}(\vx) = \mathbf{B}(\vx)^T \mathbf{w}(\vx) = \begin{pmatrix}
 (1+x_1^{l_1})\sum_{r_2\in\mathcal{R}_2} x_2^{r_2}\\
 (1+x_2^{l_2})\sum_{r_1\in\mathcal{R}_1} x_1^{r_1}
 \end{pmatrix},
 \end{equation}
 whose elements denote the locations of excited $\sigma_Z$ terms, and the two rows correspond to the two types of $\sigma_Z$ terms.
 In this case, this means that a line of $\sigma_Z[\vx^\vr(1+\bar{x}_1)]$ terms are excited along the edges of the rectangle at $r_1=0,l_1$, and a line of $\sigma_Z[\vx^\vr(1+\bar{x}_2)]$ excitations are created along the other edges at $r_2=0,l_2$.
 This is simply the domain wall excitation of the Ising model.
 
The 3D fractalized 2D Ising model is described by the Hamiltonian
 \begin{align}
H_{\text{2DIsing}}^{\text{frac}} = - \sum_{\textbf{r},s} 
 \sigma_Z[\vx^{\vr}y^s \textbf{B}(\vf(\bar{y}) \circ \bar{\vx})]
\end{align} 
for some choice of $\mathbb{F}_2$ polynomials $f_1,f_2$.

For periodic boundary conditions $L_1,L_2,L_3$ satisfying $f_1(y)^{L_1}=f_2(y)^{L_2}=1$ there are $L_3$ logical qubits encoded within the ground state manifold with pairs of independent logical operators obtained by fractalizing the logical operators in Eq.~\eqref{eq:2DQIL},
\begin{align}
\ell_{s}^X = 
\sigma_X \left[ y^s
 \sum_{\textbf{r}}
	f_1(\bar{y})^{r_1} f_2(\bar{y})^{r_2} \bar{\vx}^{\textbf{r}}
 \right] \, ,
 &&
\ell_s^Z = 
 \sigma_Z \left[ y^s
 \right] .
 \end{align}
We see that the fractalized model is also not topologically ordered as the ground space degeneracy is split by local operators. 

Suppose we now act on the ground state with a truncated $\ell^X_s$ logical operator, i.e. a fractalized $W$ operator,
$W^{\rmfrac} \equiv \sigma_X [\mathbf{w}(\vf(\vy)\circ \vx)]$.
This creates excitations located at 
 \begin{equation}
 \mathbf{E}_{W}^\rmfrac(\vx) = \begin{pmatrix}
 (1+f_1(y)^{l_1} x_1^{l_1})\sum_{r_2\in\mathcal{R}_2} f_2(y)^{r_2}x_2^{r_2}\\
 (1+f_2(y)^{l_2} x_2^{l_2})\sum_{r_1\in\mathcal{R}_1} f_1(y)^{r_1} x_1^{r_1}
 \end{pmatrix},
 \end{equation}
 which we refer to as a fractalized loop excitation.
 It is clear that such excitations have an energy cost scaling faster than linearly with the size $l_1,l_2$.  
 Thus, these are not loop-like excitations.
 They appear loop-like when projected on to the $(x_1,x_2)$ plane, but extend non-trivially into fractals in the $y$ direction.

A natural question is whether there exists \emph{any} extended topological excitations which are string-like.  
By topological, we mean that the excitation is not simply a string of locally-created excitation clusters (recall that $H^{\rmfrac}_{\mathrm{2DIsing}}$ is not topologically ordered).  
To answer this question, observe that the Hamiltonian terms obey a \emph{local relation} on each cube, where a local relation is defined to be a product of Hamiltonian terms that result in unity.
 The local relation is represented by a polynomial vector $\mathbf{l}(\bar{\vx})$
  which satisfies $\mathbf{B}(\bar{\vx})\mathbf{l}(\bar{\vx})=0$.
Going back to the original 2D Ising model, the local relation
\begin{equation}
\mathbf{l}(\bar{\vx}) = \begin{pmatrix}1+\bar{x}_2 \\ 1+\bar{x}_1\end{pmatrix} \, ,
\end{equation}
where the two columns correspond to the two Hamiltonian terms,
 tells us that $\sigma_Z[\vx^\vr \mathbf{B}(\bar{\vx}) \mathbf{l}(\bar{\vx})]=1$: a product of four Ising interaction terms around any square plaquette is trivial.
 This simply means that excitations (which are domain walls) must come in pairs around every square plaquette, and therefore form closed $\mathbb{Z}_2$ loops. 
 For the fractalized Ising model, we have the fractalized local relation
 $\mathbf{l}^\rmfrac(\bar{\vx},\bar{y}) = \mathbf{l}(\vf(\bar{y})\circ\bar{\vx})$ which implies that excitations of Hamiltonian terms are only allowed in certain configurations.  
 Meanwhile, a local cluster of excitations can always be created by acting with a single $X$ on a ground state, described by
 \begin{equation}
 \mathbf{E}_{X}(\vx,y) = \mathbf{B}^T(\vf(y)\circ\vx)=
 \begin{pmatrix}
 1+f_1(y)x_1\\
 1+f_2(y)x_2
 \end{pmatrix} \, .
 \end{equation}
 
 A topological string-like excitation, $\mathbf{E}_{\mathrm{string}}(\vx,y)$, would have to be a valid excitation satisfying $\mathbf{l}^\rmfrac(\vx,y)^T \mathbf{E}_{\mathrm{string}}(\vx,y)=0$, and also not be made up of a product of $\mathbf{E}_{X}(\vx,y)$. 
 This is equivalent to the problem of finding an $X$-type string logical operator $\ell^X = \sigma_X[\mathbf{E}_{\mathrm{string}}(\vx,y)]$ for the ``excitation Hamiltonian'' which we define in terms of the local relation and local excitation cluster,
 \begin{equation}
 H_{\mathrm{exc}} = -\sum_{\vr,s} \sigma_Z[\vx^\vr y^s \mathbf{l}^{\rmfrac}(\bar{\vx},\bar{y})] + \sigma_X[\vx^\vr y^s\mathbf{E}_{X}(\vx,y)]\\
 \end{equation}
 and is equivalent to the fractalized toric code $H^{\rmfrac}_{TC}$ ( Eq.~\ref{eq:fractc}).
 As shown in Ref.~\onlinecite{yoshida2013exotic}, this code lacks string-like logical operators iff $f_1$ and $f_2$ are algebraically unrelated, as defined in Eq.~\ref{eq:algebraicrelation}.
 Hence, for algebraically unrelated $f_{1,2}$, $H^{\rmfrac}_{2D\mathrm{Ising}}$ lacks any topological string or loop-like excitations, otherwise there exists string-like excitations (which will only be allowed to lie along a particular direction).

\subsection{3D toric code}

Next we consider fractalizing the 3D toric code, a topologically ordered model with both loop and particle excitations.
It is described by the Hamiltonian
\begin{align}
\label{eq:3DTCH}
H_{\text{3DTC}} = - \sum_{\textbf{r}} 
\sigma_X[\vx^{\textbf{r}} \textbf{A}(\vx)] + \sigma_Z[\vx^{\textbf{r}} \textbf{B}(\bar \vx)] \, ,
\end{align}
where
\begin{align} 
 \textbf{A}(\vx) =
	\begin{pmatrix}
	0 & 1+x_3 & 1+x_2  \\
	1+x_3 & 0 & 1+x_1 \\
	1+x_2 & 1+x_1 & 0
	\end{pmatrix} , &&
 \textbf{B}(\bar \vx) =
 	\begin{pmatrix}
	1+\bar{x}_1 \\
	1+\bar{x}_2  \\
	1+\bar{x}_3
	\end{pmatrix}   .
\label{eq:3DTCAB}
\end{align} 

The logical operators for the code on periodic boundary conditions correspond to conjugate string-membrane pairs given by 
\begin{align}
\ell_1^X = 
\sigma_X \left[ 
 \sum_{r_1}
 x_1^{r_1} \mathbf{e}_1
 \right]
 &&
\ell_1^Z = 
 \sigma_Z \left[ 
 \sum_{r_2,r_3}
 \bar{x}_2^{r_2}\bar{x}_3^{r_3} \mathbf{e}_1
 \right] \, ,
\end{align}
where $\mathbf{e}_1=(1,0,0)^T$ is a unit vector.
$\ell^{X/Z}_{2/3}$ are defined similarly by cyclic permutations of $1,2,3$. 
Since this model is topologically ordered, no local operators can split the ground space degeneracy. 
Truncating the string logical operator $\ell^X_i$ leads to topological quasiparticle excitations at its ends, and truncating the membrane logical operator $\ell_i^Z$ leads to a topological loop excitation along its boundary, when acting on the ground state manifold.

The 4D fractalized 3D toric code model is described by the Hamiltonian 
\begin{align}
H_{\text{3DTC}}^{\rmfrac} = - \sum_{\textbf{r},s} 
\sigma_X[\vx^{\textbf{r}} y^s \textbf{A}(\mathrm{f}( y) \circ\vx)] + \sigma_Z[\vx^{\textbf{r}}y^s \textbf{B}(\mathrm{f}(\bar y) \circ \bar \vx)] \, ,
\end{align}
for the $\textbf{A}$ and $\textbf{B}$ matrices introduced in Eq.~\eqref{eq:3DTCAB} and a choice of $\mathbb{F}_2$ polynomials $\mathrm{f}( y)=\{f_1(y),f_2(y),f_3(y)\}$.
Excitations of the $X$ term are point-like, while excitations of the $Z$ term are fractalized loops.
Similar to before, these fractalized loops resemble loops when projected to $(x_1,x_2,x_3)$, but extend non-trivially in the $y$ direction.

Following Ref.~\onlinecite{yoshida2013exotic}, a necessary and sufficient condition for the model to have no string logical operators is that the triple $f_1,f_2,f_3$ is not algebraically related. 
A triple $f_1,f_2,f_3$ is said to be algebraically related iff  an equation of the form
\begin{align}
f_1(y)^{n_1}= c~f_2(y)^{n_2}f_3(y)^{n_3} \, ,
\label{eq:3algebraicrelated}
\end{align}
holds for some finite constants $n_i,c,$ where $n_i$ are non-negative integers and not all $n_i=0$, or for some other permutation of $1,2,3$, without periodic boundary conditions.

We may also ask whether the excitations of the $Z$ terms can be string-like.
Repeating the same analysis as before, 
 we may reduce the problem to that of finding string-like logical  operators for an excitation Hamiltonian, on which the arguments of  Ref.~\onlinecite{yoshida2013exotic} can again be applied. 
The condition for the existence of string-like excitations is once  again Eq.~\ref{eq:3algebraicrelated}. 
Hence, algebraic independence of $f_1,f_2,f_3$ imply both that there  are no string-like logical operators and also no string-like $X$  excitations.

The lack of string-like $X$ excitations further implies the non-existence of a membrane $Z$-type logical operator (or some fractal structure than can be embedded onto a membrane), since if one existed then its truncation would produce a string excitation around its perimeter. 
This model still has $X$-type logical operators which are fractals that can be embedded on to a single $(x_i,y)$ plane, however.
By fractalizing the 4D toric code in the next section, we construct a 5D code which lacks any membrane logical operators altogether.

For periodic boundary conditions $L_1,L_2,L_3,L_4$ satisfying $f_i(y)^{L_i}=1$ there are $3L_4$ logical qubits encoded within the ground space and a basis of logical operators is given by 
$\ell^{X/Z,\rmfrac}_{i,s}$.
To construct an explicit example we take 
\begin{equation}
\begin{split}
f_1(y)&= 1+y+y^2 \, , \\
f_2(y)&=1+y+y^3 \, , \\
f_3(y)&=1+y^2+y^3 \, ,
\end{split}
\end{equation}
which satisfy the algebraic independence condition, as the chosen polynomials are prime, and the boundary conditions when $L_i=2^l$.

\subsection{4D toric code}

Next we consider fractalizing the 4D toric code which only supports loop-like excitations. 
The model is described by the Hamiltonian 
\begin{align}
H_{\text{4DTC}} =  - \sum_{\textbf{r}} 
\sigma_X[\vx^{\textbf{r}} \textbf{A}(\vx)] + \sigma_Z[\vx^{\textbf{r}} \textbf{B}(\bar \vx)] \, ,
\end{align}
with
\begin{align} 
\label{eq:4DTCa}
\textbf{A}(\vx) &= 
	\begin{pmatrix}
	1+x_2 & 1+x_1 & 0 & 0  \\
	1+x_4 & 0 & 0 & 1+x_1 \\
	1+x_3 & 0 & 1+x_1 & 0 \\
	0 & 1+x_3 & 1+x_2 & 0 \\
	0 & 1+x_4 & 0 & 1+x_2 \\
	0 & 0 & 1+x_4 & 1+x_3 
	\end{pmatrix}
 \, ,  
\\
\label{eq:4DTCb}
\textbf{B}(\bar \vx) &=
	\begin{pmatrix}
	0 & 0 & 1+\bar{x}_4 & 1+\bar{x}_3  \\
	0 & 1+\bar{x}_3 & 1+\bar{x}_2 & 0 \\
	0 & 1+\bar{x}_4 & 0 & 1+\bar{x}_2 \\
	1+\bar{x}_4 & 0 & 0 & 1+\bar{x}_1 \\
	1+\bar{x}_3 & 0 & 1+\bar{x}_1 & 0 \\
	1+\bar{x}_2 & 1+\bar{x}_1 & 0 & 0
	\end{pmatrix}
 \, ,
\end{align} 
where the six rows correspond to qubits living on an $(x_i,x_j)$-plaquette, with $ij=12,13,14,23,24,34$, in that order.

On periodic boundary conditions the ground space encodes six logical qubits with logical operators 
\begin{align}
\label{eq:4DTCl}
\ell_{12}^X = 
\sigma_X \left[ 
 \sum_{r_3,r_4}
 x_3^{r_3} x_4^{r_4}\mathbf{e}_{12}
 \right],
 &&
\ell_{12}^Z = 
 \sigma_Z \left[ 
 \sum_{r_1,r_2}
 \bar{x}_1^{r_1} \bar{x}_2^{r_2}\mathbf{e}_{12}
 \right],
\end{align}
where $\mathbf{e}_{12}$ is the unit vector corresponding to the qubit on the $(x_1,x_2)$ plaquette.
The logical operators $\ell_{ij}^{X/Z}$ for the remaining $ij$ are defined similarly by permuting $1,2,3,4$. 
$\ell^{X}_{ij}$ is a membrane operator acting in the plane orthogonal to $ij$, while $\ell^Z_{ij}$ acts in the plane $ij$.
This model, in fact, extends to a nontrivial finite temperature quantum phase of matter and forms a self-correcting quantum memory at finite temperature, in part due to the linear scaling of the energy barrier that needs to be overcome to implement any logical operator~\cite{Dennis2001}. 

The 5D fractalized 4D toric code is described by the following Hamiltonian 
\begin{align}
H_{\text{4DTC}}^{\rmfrac} = - \sum_{\textbf{r},s} 
\sigma_X[\vx^{\textbf{r}} y^s \textbf{A}(\mathrm{f}( y) \circ\vx)] + \sigma_Z[\vx^{\textbf{r}}y^s \textbf{B}(\mathrm{f}(\bar y) \circ \bar \vx)] \, ,
\end{align}
for the $\textbf{A}$ and $\textbf{B}$ matrices defined in Eqs.~\eqref{eq:4DTCa}~\&~\eqref{eq:4DTCb}, and some $\mathbb{F}_2$-polynomials $\mathrm{f}( y)=\{f_1(y),f_2(y),f_3(y),f_4(y)\}.$

On periodic boundary conditions $L_i,$ for $i=1 \dots 5$, such that $f_i(y)^{L_i}=1,$  for $i=1 \dots 4$,  there are $6 L_5$ encoded qubits. 
The logical operator pairs $\ell^{X/Z,\rmfrac}_{ij,s}$ are given by fractalizing Eq.~\eqref{eq:4DTCl}. 
Acting on the ground state with a truncated version of these operators, as in earlier examples, results in the creation of a fractalized loop excitation which appears loop-like when projected to a particular $(x_i,x_j)$ plane but extends non-trivially into the $y$ direction.  
Following the same argument, we find that the requirement for non-existence of string-like excitations is that $f_{1\dots 4}$ satisfy a generalized algebraic relation where  
 either 
\begin{equation}
f_1(y)^{n_1} = c~f_2(y)^{n_2} f_3(y)^{n_3} f_4(y)^{n_4}
\end{equation}
or 
\begin{equation}
f_1(y)^{n_1} f_2(y)^{n_2} = c~f_3(y)^{n_3} f_4(y)^{n_4}
\end{equation}
can be satisfied for a set of finite constants $n_i,c$ (where not all $n_i=0$) or for any
 permutation of $1,2,3,4$, without periodic boundary conditions.
 
 Hence, for four algebraically unrelated polynomials $f_i$, $H_{\mathrm{4DTC}}$ lacks any string-like excitations.
 This also implies the non-existence of any membrane logical operators,  including those with dimension less than 2 which can be embedded within a membrane. 
 This is therefore a 5D code which lacks membrane logical operators,
 a direct generalization of Type-II fracton models in 3D which are characterized by the non-existence of string-like logical operators. 
 We note that a 5D model without string-like excitations has appeared before~\cite{Haah5d}.

To find an explicit example we can take prime polynomials $f_i$. 
For instance 
\begin{align}
f_1(y)= 1+y+y^2 \, , \ \, &&
f_2(y)=1+y+y^3 \, ,\\
f_3(y) =1+y^2+y^3  \, , &&
f_4(y)=1+y+y^4 \, .
\end{align}
To achieve a more local model we can change our choice by the modification ${f_4(y)=1+y}$, at the cost of encoding fewer logical qubits for a given system size.

\subsection{Fractalized membrane excitations and beyond}

There is another toric code in 4D given by a direct generalization of the 3D toric code to include a further variable $x_4$. This model supports point-like and membrane-like excitations. Similar to the above examples it can be fractalized, leading to a model that supports ``fractalized membrane'' excitations. 
In 6D there is a toric code supporting only membrane-like excitations, which again can be fractalized. 
For an algebraically unrelated choice of fractalizing polynomials this produces a 7D model with no 3D (or lower) operators.

This follows from a dimensional reduction scheme generalizing our proof for lack of membrane logical operators. 
The presence of a 3D logical operator in $H$ implies the existence of a 2D (membrane)  excitation, which implies the existence of a 2D logical operator in its excitation Hamiltonian $H_{\mathrm{exc}}$ (as defined in Sec~\ref{sec:fracloop2d}). 
This in turn implies the existence of 1D (string-like) excitations in $H_{\mathrm{exc}}$, which implies the existence of a 1D logical operator in \emph{its} excitation Hamiltonian $H_{\mathrm{exc}^2}$.
At this point, the proof of Ref~\onlinecite{yoshida2013exotic} can be applied to show that string logical operators exist for $H_{\mathrm{exc}^2}$ iff $f_i$ are algebraically related.  
The contrapositive of this chain of implications then means that for  an algebraically unrelated choice of $f_i$, $H$ does not have any 3D logical operators. 

More generally higher dimensional toric codes exist that can support excitations of any dimensionality which can then be fractalized leading to higher dimensional extended excitation generalizations of fracton particles. 
In particular in $(2n+1)$D there is a fractalized toric code that is a generalized type-II fracton model in the sense that it supports no $2n$, or lower, dimensional logical operators. 

Fractalizing higher dimensional models requires larger sets of algebraically unrelated polynomials to construct type-II models (where the algebraic independence condition is further generalized following the form of the above generalizations). These are easy to find by simply taking all the $f_i$ to be prime polynomials of low degree, as the uniqueness of polynomial factorization over a field guarantees algebraic independence.

\section{Application: Fractal Bacon-Shor subsystem code}
\label{sec:fsc}

\subsection{Fractal Bacon-Shor code}
In this section we focus on a subsystem code obtained by fractalizing the Bacon-Shor code~\cite{bacon} and Bravyi's generalization thereof~\cite{bravyi}. We investigate the code's quantum information storage capacity which we conjecture asymptotically saturates an information storage tradeoff bound~\cite{bravyi,flammia}. 

In a subsystem code~\cite{poulin,bacon}, the full Hilbert space is structured as $\mathcal{H}=(\mathcal{H}_L\otimes \mathcal{H}_G)\oplus \mathcal{H}_E$, where $\mathcal{C}=\mathcal{H}_L\otimes\mathcal{H}_G$ is the codespace, composed of the logical and gauge subsystems, and $\mathcal{H}_E$ is an error subspace.
The information is stored in the state of the logical subsystem $\mathcal{H}_L$, while the state of the gauge subsystem can be arbitrary and is not protected. 

A subsystem code is completely specified by its gauge group $\mathcal{G}$.
From $\mathcal{G}$, the stabilizer group $\mathcal{S}$ and logical operators follow.
In this section, we explain this structure for the Bacon-Shor (BS) code~\cite{bacon} in parallel with the fractalized Bacon-Shor (FBS) code obtained by applying fractalization to the BS gauge group generators.

We take the 2D BS code defined on an $L_1\times L_2$ lattice with one qubit per site and open boundary conditions.
The gauge group $\mathcal{G}$ is generated by products of two adjacent $X$s in the same row, or $Z$s in the same column,
\begin{align}
\mathcal{G} =& \langle \{X_{\vr} X_{\vr+\mathbf{x}}, Z_{\vr} Z_{\vr-\mathbf{y}}\}\rangle
\\
=&
\left\langle
\{\sigma_X[\vx^\vr (1-x_1)],\sigma_Z[\vx^\vr(1-\bar{x}_2)]\}
\right\rangle
\equiv \langle \{A_{\vr}, B_{\vr}\}\rangle 
\, .
\nonumber
\end{align}
The Hamiltonian model based on these generators is known as the quantum compass model~\cite{Dorier2005}. 
The minus signs, although irrelevant for qubits, are included as we later generalize to $\mathbb{Z}_p$ qudits.

We apply fractalization to the gauge group generators with a pair of $1$-dimensional LCA rules, $f_1(y),f_2(y)$.
The gauge group of the FBS code is
$\mathcal{G}^{\rmfrac}=\langle \{{A}_{\vr}^{\rmfrac},{B}_{\vr}^{\rmfrac}\}\rangle$
where
\begin{equation}
\begin{split}
{A}_{\vr,s}^{\rmfrac} =& \sigma_X\left[\vx^\vr y^s (1-f_1(y) x_1) \right]\, ,\\
{B}_{\vr,s}^{\rmfrac} =& \sigma_Z\left[\vx^\vr y^s (1-f_2(\bar{y})  \bar{x}_2)\right] \, ,
\end{split}
\end{equation}
on an $L_1\times L_2\times L_3$ system, with periodic boundary conditions only along $L_3$. In the above, 
$\vr=(r_1,r_2)$ where $r_i\in R_i\equiv \{0\dots L_i-1\}$, and $s\in R_3$ with $s=L_3$ and $s=0$ identified.
The above local gauge generators define a family of Hamiltonians parameterized by the relative coupling strengths of the $X$ and $Z$ type generators, similar to the quantum compass model. Unlike stabilizer codes, such gauge code Hamiltonians may become gapless for certain values of the relative coupling strengths. 

We remark that when the three-step process of fractalization holds exactly, then the code generated by $\mathcal{G}^{\rmfrac}$ is unitarily related to the stack of $L_3$ BS codes.
Code properties that are unaffected by unitary transformations, such as the number of encoded qubits, are therefore equivalent to those of $L_3$ BS codes.
Other properties, such as the code distance which is defined below, are affected by the unitary transformation.

The stabilizer group $\mathcal{S}$ is obtained as the center of $\mathcal{G}$, $\mathcal{S}=C(\mathcal{G})\cap \mathcal{G}$, where $C(\cdot)$ is the centralizer in the Pauli group.
The codespace $\mathcal{C}$ is the simultaneous $+1$ eigenspace of all operators in $\mathcal{S}$.
For the BS code, $\mathcal{S}$ is generated by the the product of $X$ along two adjacent columns, or $Z$ along two adjacent rows,
\begin{equation}
\begin{split}
S^X_{r_1} =& \sigma_X\left[x_1^{r_1} (1-x_1) \sum_{r_2} x_2^{r_2}\right]\\
S^Z_{r_2} =& \sigma_Z\left[x_1^{L_1-1}x_2^{r_2} (1-\bar{x}_2) \sum_{r_1} \bar{x}_1^{r_1}\right]
\end{split}\,,
\end{equation}
where $S^{X}_{r_1}$ is defined for $r_1\in R_1\setminus \{L_1-1\}$, and $S^{Z}_{r_2}$ is defined for $r_2\in R_2\setminus \{0\}$ (due to the boundaries).
The generators of the stabilizer group of the FBS code,  $\mathcal{S}_{\rmfrac}$, is then 
\begin{equation}
\begin{split}
S^{X,\rmfrac}_{r_1,s} =& \sigma_X\left[ x_1^{r_1} y^s (1-f_1(y) x_1)\sum_{r_2}f_2(y)^{r_2} x_2^{r_2}\right]\, ,\\
S^{Z,\rmfrac}_{r_2,s} =& \sigma_Z\left[x_1^{L_1-1}x_2^{r_2} y^{s}(1-f_2(\bar{y}) \bar{x_2}) \sum_{r_1}f_1(\bar{y})^{r_1} {\bar{x}_1}^{r_1}\right]\, .
\end{split}\label{eq:Sfrac}
\end{equation}
for all $s\in R_3$.
These operators act on two adjacent rows or columns when projected on to $(x_1,x_2)$, like those of the BS code, but spread out non-locally into a fractal along the $y$ direction.  

The codespace is further decomposed into the logical and gauge subsystems, $\mathcal{C}=\mathcal{H}_L\otimes\mathcal{H}_G$.
The gauge operators act on the codespace as $\mathbb{1}_L\otimes G_G$.
Logical operators are those that act non-trivially in $\mathcal{H}_L$, and come in two types: bare and dressed logical operators.
Bare logical operators act as $ \ell_L\otimes\mathbb{1}_{G}$.
Dressed logical operators are products of bare logical operators and gauge operators, and therefore act on the codespace as $\ell_L\otimes G_G$.
The set of bare logical operators is defined as $L_{\mathrm{bare}} = {C}(\mathcal{S}) \setminus \mathcal{G}$.

Bare logical operators of the BS code are products of $X$ along an odd number of columns and/or $Z$ along an odd number of rows.
Let us focus on the simplest ones, consisting of products along a single row/column, which we choose to be
\begin{equation}
\begin{split}
\ell^X = \sigma_X\left[ x_1^{L_1-1} \sum_{r_2} x_2^{r_2}\right] \, ,\,
\ell^Z = \sigma_Z\left[ x_1^{L_1-1} \sum_{r_1} {\bar{x}_1}^{r_1} \right]\, ,
\end{split}
\end{equation}
and $\ell^Y=i\ell^X\ell^Z$,
from which all other logical operators can be obtained by multiplying with $\mathcal{S}$.
Analogously, the corresponding bare logical operators for the FBS code are 
\begin{equation}
\begin{split}
\ell^{X,\rmfrac}_s = \sigma_X\left[x_1^{L_1-1} y^s\sum_{r_2}f_2(y)^{r_2} x_2^{r_2}\right] \, ,\\
\ell^{Z,\rmfrac}_s = \sigma_Z\left[x_1^{L_1-1} y^s\sum_{r_1}f_1(\bar{y})^{r_1} {\bar{x}_1}^{r_1} \right]\, ,
\end{split}\label{eq:fraclog}
\end{equation}
of which there are now $L_3$ independent $X$ and $Z$ type bare logical operators.
These satisfy the commutation relation  $[[\ell^{X,\rmfrac}_{s_0} ,\ell^{Z,\rmfrac}_{s_1}]] = (-1)^{\delta_{s_0,s_1}}$, and thus generate an $L_3$-dimensional Pauli algebra on $\mathcal{H}_L$.
This code therefore protects $L_3$ logical qubits, as expected from the three-step interpretation of fractalization.
Dressed logical operators are then obtained as products of bare logical operators and gauge operators.

The code distance $d$ is defined to be the minimum support of a non-trivial dressed logical operator.
Let us take all $L_i\sim L$.
The minimum weight of bare logical operators of the FBS code scale as $d\sim L^{\mathrm{min}(d_1,d_2)}$ where $d_{1,2}$ are the Hausdorff dimensions of the fractal generated by $f_{1,2}$~\cite{yoshidaclassical}.
However, estimating the minimum weight of dressed logical operators is a difficult optimization problem in general.  
This is because it is generally possible to reduce the support of a bare logical operator by multiplying it with gauge operators.
Let us assume that $d\sim L^{\eta}$, where $\eta$ is the scaling dimension of the minimum weight dressed operator.
If $f_{1,2}$ are algebraically related, then it is straightforward to find a dressed logical operator scaling with $\eta=1$.  
However, if they are algebraically unrelated, we expect that $1<\eta\leq\mathrm{min}(d_1,d_2)\leq 2$.
The simplest example of two algebraically unrelated polynomials is $f_1(y)=1+y$ and $f_2(y)=1+y+y^2$.

We have taken periodic boundary conditions along the third dimension, which may be difficult to implement in systems with locality constraints.
This can be easily circumvented.
With open boundaries, we choose to keep all the the local gauge generators $A^\rmfrac_{\vr,s}$ and $B^\rmfrac_{\vr,s}$ with $s\in R_3$, truncating them at the edge.
With this choice of gauge generators, the logical operators are given by Eq.~\ref{eq:fraclog} with $s\in R_3$, also truncated at the edge.
As a result, logical operators near the boundary have reduced weight:
 $\ell^{Z,\rmfrac}_{0}= \prod_{r_1} Z_{(r_1,0)}$ is the minimal weight logical operator with support on only $L_1$ sites, and so $d =  L_1$.
To avoid this, we can increase the system size
$L_3\rightarrow L_3+ 2\delta L$
with 
\begin{equation}
\delta L = L_1 \mathrm{deg}(f_1) + L_2 \mathrm{deg}(f_2)
\end{equation}
where $\mathrm{deg}$ is the degree function, and use only the central $L_3$ logical qubits, $\{\ell^{X/Z,\rmfrac}_s\}_{\delta L \leq s < L_3+\delta L}$ to store information.  
The logical operators for qubits within this range are unaffected by the boundaries, and the overall scaling of the code parameters with $L$ remains the same.

\subsection{Information storage capacity and the fractal Bravyi-Bacon-Shor code}
We now discuss the information storage capacity of the fractal subsystem code.
The 2D BS code with ${n=L^2}$ physical qubits can encode $k=1$ qubits of quantum information with a code distance $d= L$.
The $[n,k,d]$ parameters of subsystem codes with a locally generated gauge group must satisfy the information storage tradeoff bound~\cite{bravyi,flammia}
\begin{equation}
k d^{\frac{1}{D-1}} \leq O(n)\label{eq:subbound}
\end{equation}
(Ref.~\onlinecite{bravyi} only derived the bound in $D=2$, but the proof can be extended to higher dimensions as written down in Ref.~\onlinecite{flammia}).
Bravyi's generalization~\cite{bravyi} of the BS code, henceforth referred to as the Bravyi-Bacon-Shor (BBS) code, saturates this bound in 2D.

The 2D BBS code is parameterized by a binary matrix $K_{\vr}$, which specifies a set of sites $K_\vr=0$ to be removed from the 2D lattice.
The gauge generators are modified accordingly to skip over the removed sites, i.e. $X_{\vr} X_{\vr+n\mathbf{x}}$, where $\vr$,$\vr+n\mathbf{x}$ are two nearest unremoved sites along that row.
Although this may induce long range couplings, additional auxiliary qubits may be introduced to ensure all generators are local~\cite{bravyi}.
The end result is that logical operators from different rows or columns can no longer generically be related to one another by $\mathcal{S}$, and the number of logical qubits is given by $k=\mathrm{rank}(K)$.
Furthermore, it is possible~\cite{bravyi} to find a family of matrices $K_\vr$ such that
$k\sim L$, while maintaining $n\sim L^2$ and $d\sim L$.
Thus, $[n,k,d]\sim[L^2,L,L]$ saturates the 2D bound in Eq.~\ref{eq:subbound}, $k d \sim L^2 \leq \mathcal{O}(L^2)$.
In 3D, however, the current best known subsystem code does not quite saturate the scaling in the tradeoff bound~\cite{baconflammia}.

The FBS code has $n=L^3$, $k=L$, and $d\sim L^{\eta}$ for some $1\leq \eta \leq 2$ as discussed previously. 
This is still far from saturating the 3D tradeoff bound, $k\sqrt{d}\sim L^{1+\eta}\leq \mathcal{O}(L^3)$.

To further improve the quantum information storage capacity, we may apply fractalization to the BBS code, resulting in the fractal BBS (FBBS) code.
Removing a single site $K_\vr=0$ from the original model corresponds to removal of a whole line of sites $\{(\vr,s)\}_{s}$ from the fractal model.  
The number of logical qubits is improved to $k=L  \,\mathrm{rank}(K)$.
Again, there is a choice of binary matrices such that the parameters scale as $[n,k,d]\sim [L^3,L^2,L^{\eta}]$ (for a potentially different $\eta$).
 Thus, the FBBS code nearly saturates the tradeoff bound depending on the value of $\eta$,
 $k\sqrt{d}\sim L^{2+\eta}\leq O(L^3)$.

Next, we may further generalize beyond qubits to qudits of prime dimension $p$.  
In the limit of large $p$, the Hausdorff dimensions of the generated fractals approach $d_{1,2}\rightarrow 2$ (this was used by Yoshida~\cite{yoshidaclassical} to asymptotically saturate the information storage bound in a classical spin system).
We conjecture that in the limit $p\rightarrow\infty$, for $f_1,f_2$ algebraically unrelated polynomials, the code distance scaling parameter also approaches $\eta\rightarrow 2$.  
The FBBS code is therefore conjectured to asymptotically saturate the 3D information storage bound $k\sqrt{d}\sim n$ in the limit of large $p$.

We may also generalize to higher dimension.  
Suppose we wish to apply fractalization to a code from ${D\rightarrow D+m}$ dimensions.
Let us start with an $[n,k,d]$ subsystem code on an $n=L^D$ system, and apply fractalization via the three-step process as described in Sec~\ref{sec:interp}.
The first step is to construct a layered system, in which the code now has parameters $[L^m n, L^m k, d]$.
The second step is a different choice of gauge generators, which does not affect any parameters of the code.
Finally, the third step is a non-local unitary transformation.  
This does not affect $n$ or $k$, but does affect the distance, which now scales as  $d\sim L^{\eta}$, where $1\leq \eta \leq m+1$.
Thus, starting with the 2D BBS code, the $2+m$ dimensional FBBS code has parameters $[n,k,d]\sim [L^{m+2},L^{m+1},L^{1\leq \eta \leq m+1}]$.
The analogue of our conjecture in higher dimensions would be that, for certain choices of LCA rules and $\mathbb{Z}_p$ qudits, $\eta\rightarrow m+1$ in the limit $p\rightarrow\infty$.  
In this case, the tradeoff bound 
\begin{equation}
k d^{\frac{1}{m+1}}\sim L^{m+1+\eta/(m+1)}\xrightarrow[\eta\rightarrow m+1]{p\rightarrow\infty} L^{m+2} \leq O(L^{m+2})
\end{equation}
is saturated in the large $p$ limit in any dimension $m+2$.

\section{Conclusions}
\label{sec:conclusion}

We have introduced the fractalization procedure, by which certain spin models can be extended into higher dimensional fractal models.
Many exotic fractal models, such as type-II fracton models or fractal SPTs, may be understood as fractalized versions of familiar spin models, such as the toric code or the cluster model, as we have discussed in detail with many examples.
We provided an interpretation of fractalization in terms of a three-step process, which elucidates how the properties of the fractal model are inherited from those of the original model.
Fractalizing models with loop-like or higher dimensional excitations, for example, leads to exotic extended fractal excitations.

We have introduced the fractal Bacon-Shor subsystem code which was, after applying Bravyi's generalization~\cite{bravyi} and in the limit of large on-site Hilbert space dimension~\cite{yoshidaclassical}, conjectured to saturate the information storage tradeoff bound in 3D.  
A more detailed analysis of the code properties of the fractal code is important, but beyond the scope of this current paper.  
This would include a numerical estimation of the code distance, especially in the limit of large $p$, as well as finding and implementing error correction algorithms. Due to the presence of non-local stabilizers, similar to the BS code, we expect that the fractalized BS code does not give rise to a positive threshold for local noise, without further modifications. 

There are still many open directions regarding the fractalization procedure.  
Currently, fractalization can only be applied to purely $X$ or $Z$ operators  while preserving commutativity; 
it would be interesting if a generalization beyond such operators is possible.  
An intriguing potential application would then be the fractalization of Kitaev's honeycomb  model~\cite{kitaev2006anyons} in the non-Abelian Ising phase to possibly obtain a non-Abelian type-II fracton model. 
Other possibilities would be generalizing beyond translation-invariant LCA, perhaps to quantum cellular automata.  
These would allow fractalization to be applied to more complicated spin models.
Unfortunately, the present formalism in this paper does not readily admit such generalizations.
Another question is whether there exists an interpretation, similar to the three-step interpretation, of fractalization with higher-order LCA.  
Such an interpretation would greatly improve our understanding of type-II fracton phases described by higher-order LCA, such as Haah's cubic code.

There are also many more CSS codes to which fractalization can be applied, particularly in higher dimensions. 
In this work we considered fractalizing 3D and 4D toric codes, the latter lead to 5D models with no membrane operators that are, in some sense, a generalization of the type-II fracton models that exist in 3D.  
Beyond this, each other $D$ dimensional toric code or type-I fracton model defines a family of $D+m$ dimensional fracton models with fractal logical operators and excitations beyond the examples covered in the current work.

\acknowledgements
TD acknowledges support from the Charlotte Elizabeth Procter Fellowship at Princeton University. 
DW acknowledges support from the Simons foundation. 

\appendix
\section{CX circuit for arbitrary M}\label{app:cx}
In this section, we present an algorithm for constructing a CX circuit $U$ that implements the transformation
\begin{equation}
U\sigma_X[\mathbf{v}]U^{\dagger} =\sigma_X[\mathbf{M}\mathbf{v}], \;\;\;\;
U\sigma_Z[\mathbf{w}]U^\dagger = \sigma_Z[\mathbf{M}^{-1,T}\mathbf{w}].
\end{equation}
for an arbitrary invertible binary matrix $\mathbf{M}\in\mathbb{F}_2^{L\times L}$ with $M_{ii}=1$.
We will say that $U$ is the unitary corresponding to the matrix operation $\matr{M}$.

A single $\mathrm{CX}_{ij}$ gate ($i\neq j$), defined as
\begin{equation}
\mathrm{CX}_{ij} = (Z_i-1)/2 + (Z_i+1) X_j / 2\,,
\end{equation}
corresponds to the matrix operation 
\begin{equation}
\mathbf{T}_{ij} = \mathbb{1} + \mathbf{e}_i \mathbf{e}_j^T
\end{equation}
where $\mathbf{e}_i$ is the $i$th unit vector.
Thus, the problem of finding the CX circuit for $\mathbf{M}$ is equivalent to finding a sequence of pairs $(i_n,j_n)$, $n=1\dots N$, that satisfies
\begin{equation}
\mathbf{M} = \mathbf{T}_{i_1 j_1}\dots \mathbf{T}_{i_N j_N}
\label{eq:MT}
\end{equation}
corresponding to the CX circuit 
\begin{equation}
U=\mathrm{CX}_{i_1 j_1}\dots \mathrm{CX}_{i_N j_N}.
\end{equation}

We can construct $\matr{M}$ iteratively by noticing that
\begin{equation}
(\matr{T}_{i j} \matr{M})_{kl} = M_{kl} + \delta_{ik}M_{jl} 
\end{equation}
corresponds to a row action on $\matr{M}$ where the $j$th row of $\matr{M}$ is added to the $i$th row.  
As $\matr{M}$ is invertible with unit diagonal, a series of such row operations can be used to transform $\matr{M}$ into the identity matrix.
The sequence of row operations $(i_n,j_n)$ may be generated via the following algorithm,
\begin{itemize}
\item[] Start with $\matr{M}$ and $n=1$.
\item[] For each $j=1\dots L$,
\begin{itemize}
\item[] For each $i=1\dots L$, except $i=j$,
\begin{itemize}
\item[] If $M_{ij}=1$, set $(i_n,j_n)=(i,j)$, 
\begin{equation*}
\matr{M}\rightarrow \matr{M}^\prime =\matr{T}_{ij}\matr{M},  
\end{equation*}
\,\, and then increment $n\rightarrow n+1$.
\end{itemize}
\end{itemize}
\end{itemize}

By the end of this algorithm, $\matr{M}$ will have been transformed into the identity matrix.
Thus, we have generate a list of pairs $(i_n,j_n)$ such that
\begin{equation}
\matr{T}_{i_N,j_N}\dots \matr{T}_{i_1,j_1}\matr{M} = \mathbb{1}
\end{equation}
Since $\matr{T}_{ij}^2 = \mathbb{1}$, this can be rewritten
\begin{equation}
\matr{M} = \matr{T}_{i_1,j_1}\dots \matr{T}_{i_N,j_N}
\end{equation}
which matches Eq.~\ref{eq:MT} and we have therefore found a CX circuit corresponding to $\matr{M}$.


\bibliographystyle{unsrtnat}
\bibliography{refs}




\end{document}